\documentclass[12pt]{article}
\usepackage{epsf,amsmath, amsfonts}
\usepackage{epsfig,mathtools}
\usepackage{color,graphicx}% Include figure files
\usepackage{dcolumn}% Align table columns on decimal point
\usepackage{bm}% bold math
\usepackage{epsfig,latexsym}
\usepackage{lineno,newtxtext,nicefrac,bigints}
\usepackage{dashrule,hyperref}
\usepackage{mathptmx}      % use Times fonts if available on your TeX system

\usepackage[skip=0pt]{caption}
\usepackage[skip=0pt]{subcaption}
\newcommand{\bseq}{\begin{subequations}}
\newcommand{\eseq}{\end{subequations}}
\newcommand{\beq}{\begin{equation}}
\newcommand{\eeq}{\end{equation}}
\newcommand{\bef}{\begin{figure}}
\newcommand{\eef}{\end{figure}}

\hypersetup{
    colorlinks=true,
    urlcolor=red,
}

\let\OLDthebibliography\thebibliography
\renewcommand\thebibliography[1]{
  \OLDthebibliography{#1}
  \setlength{\parskip}{0pt}
  \setlength{\itemsep}{1pt plus 0.3ex}
}

\begin{document}

%\preprint{AIP/123-QED}

\title{Spatiotemporal linear stability of viscoelastic free shear flows: non-affine response regime}
\author{ {\bf D. Bansal$\dagger$}, {\bf D. Ghosh$\dagger$} and {\bf S. Sircar$\dagger, \ddagger$}
\\{$\dagger$ \small Department of Mathematics, IIIT Delhi, India 110020} \\{\small $\ddagger$ Corresponding Author (email: sarthok@iiitd.ac.in)}}
\maketitle
\date{ }% It is always \today, today,
            
\begin{abstract}
We provide a detailed comparison of the two-dimensional, temporal and the spatiotemporal linearized analyses of  the viscoelastic free shear flows in the limit of low to moderate Reynolds number and Elasticity number obeying four different types of stress-strain constitutive equations: Oldroyd-B, Upper Convected Maxwell, Johnson-Segalman (JS) and linear Phan-Thien Tanner (PTT). The resulting fourth-order Orr-Sommerfeld Equation is transformed into a set of six auxiliary equations that are numerically integrated via the Compound Matrix Method. The temporal stability analysis suggest (a) elastic stabilization at higher values of elasticity number (shown previously in the dilute regime [S. Sircar and D. Bansal, ``Spatiotemporal linear stability of viscoelastic free shear flows: Dilute regime'', Phys. Fluids {\bf 31}, 084104 (2019)]), (b) a non-monotonic instability pattern at low to intermediate values of elasticity number for the JS as well as the PTT model. To comprehend the effect of elasticity, Reynolds number and viscosity on the temporal stability curves of the PTT model, we consider a fourth parameter, the centerline shear rate, $\zeta_c$. The `JS behaviour' is recovered below a critical value of $\zeta_c$ and above this critical value the PTT base stresses (relative to the JS model) is attenuated thereby explaining the stabilizing influence of elasticity. The Briggs idea of analytic continuation is deployed to classify regions of temporal stability, absolute and convective instabilities, as well as evanescent modes, and the results are compared with previously conducted experiments for Newtonian as well as viscoelastic flows past a cylinder. The phase diagrams reveal the two familiar regions of inertial turbulence modified by elasticity and elastic turbulence as well as (a recently substantiated) region of elastoinertial turbulence and the unfamiliar temporally stable region for intermediate values of Reynolds and Elasticity number. 
\end{abstract}
\maketitle


%%%%%%%%%%%%%%%%%%%%%%%%%%%%%%%%%%%%%%%
\section{\label{sec:level1}Introduction} \label{sec:intro}
The understanding of the hydrodynamic stability and the flow transition in free shear flows of viscoelastic liquids has continued to receive prolonged interest due to its practical applications in microfluidic mixing~\cite{Squires2005}, viscoelastic stabilization via polymer addition~\cite{White2008} and in shearing flows of viscoelastic biofluids, like mucus~\cite{Sircar2016DCDS}, cartilage~\cite{Sircar2015} and adhesion-fragmentation transition in cells~\cite{Sircar2013,Sircar2014,Sircar2016JOMB,Sircar2016EPEJ}. Most investigations of low to moderate Reynolds number ($Re$) have either focused on the linearized studies~\cite{Ray2014}, experimental analysis~\cite{Groisman1998} or the full Direct Numerical Simulations (DNS)~\cite{Yu2004} of the affine response of dilute polymeric liquids. However, in  semi-dilute or moderately concentrated polymeric liquids, instabilities may arise due to flow induced inhomogeneities, and an improved understanding of the transition pathway of the non-affine or the non monotonic flow response in (but not limited to) strong elongational flow, is expedient.

Using the local spatiotemporal stability analysis where the base flow is assumed to be locally parallel~\cite{Huerre1985}, we search for absolute instabilities (perturbations which grow exponentially in time at the point of excitation), convective instabilities (disturbances which are swept downstream from the source and decay at any fixed position in space) and the evanescent modes (or false modes)~\cite{Patne2017}. The local approach is best-suited for moderate to high $Re$, especially for slowly spreading shear flows. Flow-spreading effects can be fully captured with global methods though the computational cost is substantially higher. Juniper \emph{et al.}~\cite{Juniper2011} discusses the complementary nature of local and global methods in their study of the stability of confined Newtonian wakes.

The concepts of absolute and convective instabilities are well established to study the evolution of impulse disturbances in Newtonian flows. Absolute instability was earlier experimentally verified by Shoji \emph{et. al}~\cite{Shoji2020} in liquid jets. There is also an extensive literature on Newtonian wakes and mixing layers, including the blunt body experiments listed by Oertel~\cite{Oertel1990}, the linear analysis by Chomaz~\cite{Delbende1998} and Healy~\cite{Healy2009} and the DNS studies by Pier~\cite{Pier2008}. Overall, these studies found a destabilizing effect of finite boundaries and an existence as well as a transition to absolute instability in the near wake region.

The spatiotemporal analysis of viscoelastic free shear layers are more recent and scarce. 
An early experimental study by Vihinen \emph{et. al}~\cite{Vihinen1997} reported absolute instability in viscoelastic liquid jets. Pipe reported a stabilizing effect of polymer addition in his experiments on viscoelastic cylindrical wakes, which is counteracted by shear thinning and a transition from convective to absolute instability at higher polymer concentrations~\cite{Pipe2005}. In contrast, the linear analysis of dilute mixing layers~\cite{Ray2014} and dilute jets~\cite{Ray2015, Alhushaybari2019, Alhushaybari2020} relay a significant range of parameters where viscoelasticity was found to be destabilizing. A recent DNS study of jets~\cite{Guimaraes2020} necessitates the use of an extra (convective) timescale to characterize the memory fading property of viscoelastic fluids.

We have limited our focus on the linear spatiotemporal analyses of viscoelastic free shear layer, mainly within the regime of non-affine / non monotonic flow response. We characterize the fluid flows as being absolutely or convectively unstable and utilize the method of spatiotemporal analysis by progressive moving of the isocontours in the complex frequency and wavenumber plane, as proposed by Kupfer~\cite{Kupfer1987}. The next section describes the model of the viscoelastic free shear layer flow coupled with four constitutive relations for the extra elastic stress tensor for simultaneous comparison of results: Oldroyd-B, Upper Convected Maxwell (UCM), Johnson-Segalman (JS) at the slip parameter value, $a=0.5$, and the linear Phan-Thien and Tanner (PTT) model at $a=0.5$ and the elongation parameter explored within the range $\varepsilon \in [0,  0.99]$, although a detailed study is predominantly conducted at $\varepsilon=0.5$ (\S \ref{subsec:mm}) as well as the details of the linear stability analysis leading upto the fourth order Orr-Sommerfeld equation (OSE) (\S \ref{subsec:lsa}). \S \ref{sec:CMM} introduces the Compound Matrix Method (CMM) to numerically solve the resultant system of stiff ordinary differential equations (ODE) emerging from the OSE, rewritten in terms of the auxiliary variables. \S \ref{sec:results} showcases the simulation results including numerical method validation (\S \ref{subsec:mv}), the temporal (\S \ref{subsec:tsa}) and the spatiotemporal stability analyses (\S \ref{subsec:stsa}), followed with a brief discussion on the implication of these results as well as the focus of our future direction (\S \ref{sec:conclusion}). The appendix (\S \ref{sec:appendix}) lists all the coefficients of the OSE.

%\vspace{-1cm}
%%%%%%%%%%%%%%%%%%%%%%%%%%%%%%%%%%%%%%%
\section{\label{sec:level1}Mathematical model and linear stability analysis}\label{sec:mm_lsa}
Unlike Newtonian solvents, the transition to instability in polymeric liquids depends on the details of the equations relating stress to the shear rate. The linear PTT model properly describes the non-affine behavior of polymeric liquids composed of a low to moderate concentration of high molecular weight polymer in a very viscous Newtonian solvent at moderate shear rates, including the effect of shear thinning, non-zero normal stress coefficient and stress overshoot in transient flows~\cite{Yu2004}. Further, this model predicts finite stress at finite strain rates for strong elongation flow, a feature which is achieved by constraining the length of the polymer chain to a maximum allowable length. The numerical solution obtained by linearizing the Navier Stokes along with the extra stress constitutive equation is reliable in describing the characteristics of the initial stages of the mixing layer transition~\cite{Bird1987Vol1}. Hence, the linear stability results can be utilized as the initial-boundary conditions for Direct Numerical Simulation~\cite{Yu2004} or the Large Eddy Simulation of polymeric liquids~\cite{Steinberg2018} in order to further study the flow evolving process downstream.

%\vspace{-0.5cm} 
%%%%%%%%%%%%%%%%%%%%%%%%%%%%%%%
\subsection{\label{sec:level2}Mathematical model}\label{subsec:mm}
The continuity and the momentum equations for an incompressible flow in a free shear flow configuration are as follows,
\vspace{-0.5cm} 
\bseq
\begin{align}
\nabla\cdot {\bf v} &= 0, \label{eqn:MainModel1}\\
\rho \frac{D {\bf v}}{D t} &= -\nabla p + \nabla \cdot \tau, \label{eqn:MainModel2}
\end{align}
\label{eqn:MainModel} 
\eseq
where ${\bf v}$ is the velocity vector, $\rho$ the density, $p$ the isotropic pressure and $\tau$ the extra stress tensor, which is written as the sum of the viscous, Newtonian stress, $\tau^s$ (= $\eta_s {\bf D}$, where ${\bf D}=\nabla {\bf v} + \nabla {\bf v}^{\text T}$ is the rate of strain tensor) and the elastic stress, $\tau^p$ (= $\eta_p {\bf A}$). $\eta_s, \eta_p$ are the solvent viscosity and the polymeric contribution to the shear viscosity, respectively. Introducing the parameters, $\eta (= \eta_s + \eta_p)$ and $\nu (= \eta_p/\eta)$, representing the total viscosity and the elastic contribution to the total viscosity, respectively, the extra stress tensor can be rephrased as follows,
\beq
\tau = \eta [(1-\nu) {\bf D} + \nu{\bf A}].
\label{eqn:Stress}
\eeq
The tensor ${\bf A}$ satisfies the linear PTT equation~\cite{Yu2004},
\beq
\left(1 + \varepsilon \lambda \text{tr}{\bf A}\right) {\bf A} + \lambda \frac{D A}{D t} = {\bf D},
\label{eqn:JS1}
\eeq
where the Gordon-Schowalter (GS) convected derivative, {$\displaystyle \frac{D A}{D t}$}, is,
\beq
\frac{D A}{D t} = \frac{\partial {\bf A}}{\partial t} + {\bf v} \cdot \nabla {\bf A} -  \frac{1}{2}({\bf W}^{\text T} + a{\bf D})\cdot {\bf A} - \frac{1}{2}{\bf A} \cdot ({\bf W} + a{\bf D}). 
\label{eqn:JS2}
\eeq
${\bf W}=(\nabla {\bf v}\!\!-\!\!\nabla {\bf v}^{\text T})$ is the vorticity tensor, $\lambda$ is the polymer relaxation time, $a$ is a slip parameter characterizing the non-affine motion of the chains and $\varepsilon \in [0, 1]$ is a dimensionless parameter describing the maximum elongation of the polymer chains. For the parameters, $\varepsilon=0, a = 1$, the motion is affine and the PTT model reduces to the Oldroyd-B model. The Oldroyd-B model describes well the behavior of dilute polymeric liquids composed of a low concentration of high molecular weight polymer in a very viscous Newtonian solvent at moderate shear rates. The model predicts no shear thinning, a constant first normal stress coefficient and a zero second normal stress coefficient. But this constitutive relation fails in many circumstances, e.~g., it fails to predict the physical value of the viscosity in extensional flow when the relaxation time of the polymer times the extension rate exceeds 0.5 (or when the elongational flow is strong enough to drive the two sides of the polymer dumbbell infinitely far apart from one another)~\cite{Bird1987Vol1}.

An particular case of the Oldroyd-B model is the UCM model ($\varepsilon=0, a=1, \nu=1$), a case where the extra stress tensor is purely elastic. A UCM liquid predicts a quadratic shear rate dependence of the first normal stress difference and zero second normal stress difference (which is a realistic behavior of polymer melts at moderated shear rates~\cite{Bird1987Vol1}), but a constant shear viscosity (or no shear thinning behavior).

The JS model (illustrated in detail at a fixed value of slip parameter, $a=0.5$, and $\varepsilon=0$) allows for a non-monotonic relationship between the shear stress and rate of shear in a simple shear flow, consequently explaining the `spurt' phenomena or a dramatic increase of the volumetric flow rate (equivalently a spike in the strain rate) at a critical stress which is independent of molecular weight of the polymer~\cite{Malkus1991}.

Finally, the linear PTT model (illustrated via the model parameters $a=0.5, \varepsilon \in [0, 0.99]$) is based on finitely extensible springs; it does reproduce
shear-thinning and generally captures the instability transition more accurately than the Oldroyd-B as well as the JS model~\cite{Yu2004}. In particular, this model is useful in obtaining a closer fit to the real world normal stress differences including the study of contraction and re-entrant corner flows~\cite{Evans2010}.

%\vspace{-0.75cm}
%%%%%%%%%%%%%%%%%%%%%%%%%%%%%%%%%%%%%%%%%%%
\subsection{\label{sec:level2}Linear Stability Analysis}\label{subsec:lsa}
Consider the free-stream velocity in a frame moving with the average flow velocity, i. e., $U$ ($= \frac{1}{2}$($\mathcal{U}_1$ - $\mathcal{U}_2$), where $\mathcal{U}_1$($\mathcal{U}_2$) is the free-stream velocity of the upper(lower) flow), and the momentum thickness, $\delta$~\cite{Azaiez1994}. Utilizing $U, \delta$ as the reference velocity and the length scale, respectively, we nondimensionalize equation~\eqref{eqn:MainModel} and introduce the the dimensionless numbers, $Re = \nicefrac{\rho \delta U}{\eta}$ and the Weissenberg number, $We = \nicefrac{\lambda U}{\delta}$. Assuming that the mean flow is two-dimensional (with $x$ and $y$ being the streamwise parallel and transverse directions in space, respectively) and quasi-parallel with its variation entirely in the transverse direction, i.~e., 
%\vspace{-0.25cm}
\beq
U(y) = \tanh(y), \,\, \Omega(y) = \tanh^2(y)-1, \,\, \Psi(y) = \log(\cosh(y)),
\label{eqn:Meanflow}
\eeq
%\vspace{-0.25cm}
where $U(y), \Omega(y), \Psi(y)$ are the dimensionless streamwise parallel mean velocity (with zero transverse component), mean vorticity and the associated streamfunction, respectively. Further, assuming that the mean flow supports a two-dimensional disturbance field, the streamfunction and the extra stress tensor are represented by the base state profile $(\Psi(y), \tau_0(y))$ plus a small perturbation, which is Fourier transformed in $x$ and $t$ as follows,
\begin{align}
\psi(x,y,t) &= \Psi(y) + \phi(y) e^{{\it i}(\alpha x - \omega t)}, \nonumber \\
\tau(x,y,t) &= \tau_0(y) + \varphi(y) e^{{\it i}(\alpha x - \omega t)}, 
\label{eqn:disturbance}
\end{align}
where $\phi(y), \varphi(y)$ are the transverse perturbations in the streamfunction and the extra stress tensor and $\alpha, \omega$ are the complex wavenumber and angular frequency, respectively. We note that equation~\eqref{eqn:Meanflow} is a solution of the momentum equations for incompressible flow provided there is a dimensionless body force term on the right-hand-side of equation~\eqref{eqn:MainModel2}~\cite{Sircar2019}. We rephrase equations~(\ref{eqn:MainModel}-\ref{eqn:JS2}) in the streamfunction-vorticity formulation and avail equation~\eqref{eqn:disturbance} to arrive at the equation governing the perturbation of the streamfunction, given by the fourth order OSE~\cite{Azaiez1994}.
\beq
\left\{\!\!i \left[\!(\alpha U\!\!-\!\!\omega)(D^2 \!\!-\!\! \alpha^2) \!\!-\!\! \alpha D^2U \!\right] \!\!-\!\! \frac{1-\nu}{Re} (D^2 \!\!-\!\! \alpha^2)^2 \!\!\right\} \phi \!\! = \frac{\nu}{\mathcal{F} Re} \!\!\sum_{n=0}^{4} \!\!c_n D^n \phi.
\label{eqn:OSE}
\eeq
where $D^n(\cdot)$ denote the n$^{\text th}$ derivative of any variable function with respect to $y$ and the coefficients $c_i$'s (equations~\eqref{eqn:OSE_ci}) are listed in \S \ref{sec:appendix}. $D = \nicefrac{d}{dy}$, $\mathcal{F} = 4 d^3_0 \left[ a S_1 S_2 + S_1 S_3 + (1-a^2)S^2_1\right]^2\left[ S_3 + (1+a)S_1 \right]^2$ (refer \S \ref{sec:appendix}). Following the usual Newtonian development~\cite{Azaiez1994}, we assume that $\phi = \phi_r + {\it i} \phi_i$ (the subscript $r/i$ from this point onwards will denote the real and the imaginary components) and $\phi_r$ and $\phi_i$ (both real valued functions) are even and odd functions of $y$ respectively. Thereby, we restrict the domain of integration to the upper half of the flow such that the boundary conditions at $y = 0$ (or the so called centerline conditions), are altered as follows,
\bseq
\begin{align}
&\phi_i = \phi''_i = 0 \label{eqn:CL_AS}\\
&\phi'_r = \phi'''_r= 0.\label{eqn:CL_S}
\end{align}
\label{eqn:CL} 
\eseq
The far stream boundary conditions (or the conditions at $y \rightarrow \infty$) is given by
\beq
\phi_{r/i} = \phi'_{r/i} = 0. \label{eqn:FS}
\eeq
In \S~\ref{sec:CMM}, we detail a solution procedure for solving the eigenvalue problem~(\ref{eqn:OSE}-\ref{eqn:FS}).
%
%%%%%%%%%%%%%%%%%%%%%%%%%%%%%%%%%%%%%%%%%%%%%%
\section{\label{sec:level1}Solution to the eigenvalue problem} \label{sec:CMM}
The primary step in evaluating the eigenvalues ($\alpha, \omega$) and the eigenfunction, $\phi$, describing the disturbance field (satisfying equation~\eqref{eqn:OSE}) is to probe the ramification of the far stream boundary conditions~\eqref{eqn:FS} by using the mean flow information at far stream, i.~e., $\lim y \rightarrow \infty: U(y)=1, U''(y)=0$, which simplifies the OSE~\eqref{eqn:OSE} to the following constant coefficient ODE~\cite{TKS2012},
\beq
\phi^{(4)} - 2\alpha^2 \phi'' + \alpha^4 \phi = \frac{{\it i} Re S_\infty}{\nu + (1-\nu) S_\infty} (\alpha - \omega)(\phi'' - \alpha^2 \phi),
\label{eqn:OSE-CC}
\eeq
where $S_\infty=1+We(\alpha - \omega)$. The solution to equation~\eqref{eqn:OSE-CC} can be derived by setting $\phi=e^{\lambda y}$, such that one gets the characteristic roots as $\lambda_{1,2} = \mp \alpha$ and $\lambda_{3,4} = \mp Q$, where $Q\!\!=\!\!{\displaystyle \left[ \alpha^2 \!\!+\!\! \frac{{\it i} Re S_\infty(\alpha \!\!-\!\! \omega)}{\nu \!\!+\!\! (1-\nu) S_\infty} \!\right]^{\frac{1}{2}}}$. The fourth order OSE~\eqref{eqn:OSE} has four fundamental solutions, i.~e. $\{\phi_i\}^4_{i=1}$, whose asymptotic variation with $y \rightarrow \infty$ is: $\phi_{1,2} \sim e^{\mp\alpha y}$; $\phi_{3,4} \sim e^{\mp Q y}$. Then the general solution which satisfies the far stream conditions~\eqref{eqn:FS} for real $(\alpha, Q) > 0$, is of the form
\beq
\phi = a_1 \phi_1 + a_3 \phi_3.
\label{eqn:general}
\eeq
Equation~\eqref{eqn:general} admits a non-trivial solution of the OSE, satisfying the far stream condition~\eqref{eqn:FS} and the centerline conditions~(\ref{eqn:CL_AS}, \ref{eqn:CL_S}) if and only if the determinant of the associated matrix of the linear algebraic system given by equation~\eqref{eqn:general}, vanishes at $y=0$, or
\beq
\left(\phi_1 \phi''_3 - \phi''_1 \phi_3\right)|_{y=0} = 0,
\label{eqn:OSE_DRP_AS}
\eeq
for the odd component (equation~\eqref{eqn:CL_AS}) and 
\beq
\left(\phi'_1 \phi'''_3 - \phi'''_1 \phi'_3\right)|_{y=0} = 0,
\label{eqn:OSE_DRP_S}
\eeq
for the even component (equation~\eqref{eqn:CL_S}), respectively. Equations~(\ref{eqn:OSE_DRP_AS}, \ref{eqn:OSE_DRP_S}) serve as the dispersion relation of the problem and will be solved simultaneously. The stiffness of the OSE~\eqref{eqn:OSE} (e.~g., in the case of far stream eigenmodes in the limit $Re \rightarrow \infty$, we see that $|Q| \gg |\alpha|$, leading to an immense contrast between the two sets of characteristic roots of equation~\eqref{eqn:OSE-CC}) thereby causing the solution components corresponding to the different fundamental solutions to lose linear independence. This source of parasitic error growth necessitates the use of CMM~\cite{Ng1985}, where one works with a set of following auxiliary variables,
\begin{align}
&y_1 = \phi_1 \phi'_3 - \phi_3 \phi'_1, \quad y_2 = \phi_1 \phi''_3 - \phi_3 \phi''_1, \quad y_3 = \phi_1 \phi'''_3 - \phi_3 \phi'''_1, \nonumber \\
&y_4 = \phi'_1 \phi''_3 - \phi''_1 \phi'_3, \quad y_5 = \phi'_1 \phi'''_3 - \phi'''_1 \phi'_3, \quad y_6 = \phi''_1 \phi'''_3 - \phi'''_1 \phi''_3,
\label{eqn:AuxiliaryVar}
\end{align}
satisfying the initial value problem (IVP)~\cite{Sircar2019},
\begin{align}
y'_1 \!\!&= \!\!y_2, \nonumber\\
y'_2 \!\!&= \!\!y_3 + y_4, \nonumber\\
y'_3 \!\!&= \!\!y_5\!\!-\!\!\left[\frac{(1\!\!-\!\!\nu)(c_1 y_1\!\!+\!\!c_2 y_2\!\!+\!\!c_3 y_3)\!\!-\!\!\mathcal{F}({\it i}Re(\alpha U\!\!-\!\!\omega\!\!+\!\!2\alpha^2\nu)y_2}{(1\!-\!\nu)c_4\!+\!\mathcal{F}\nu}\right], \nonumber\\
y'_4 \!\!&= \!\!y_5, \nonumber\\
y'_5 \!\!&= \!\!y_6\!\!+\!\!\left[\frac{(1\!\!-\!\!\nu)(c_0 y_1\!\!-\!\!c_2 y_4\!\!-\!\!c_3 y_5)\!\!+\!\!\mathcal{F}({\it i}Re(\alpha U\!\!-\!\!\omega)\!\!+\!\!2\alpha^2\nu)y_4}{(1\!-\!\nu)c_4\!+\!\mathcal{F}\nu} \right.\nonumber\\
&\left. + \frac{\mathcal{F}({\it i}Re[ (\alpha U\!\!-\!\!\omega)\alpha^2\!\!+\!\!\alpha U'']\!\!+\!\!\alpha^4\nu)y_1}{(1\!-\!\nu)c_4\!+\!\mathcal{F}\nu}\right], \nonumber\\
y'_6 \!\!&= \!\!\frac{(1\!\!-\!\!\nu)(c_0 y_2\!\!+\!\!c_1 y_4\!\!-\!\!c_3 y_6)\!\!+\!\!\mathcal{F}\!({\it i}Re[ (\alpha U\!\!-\!\omega)\alpha^2\!\!+\!\!\alpha U''])y_2}{(1\!-\!\nu)c_4\!+\!\mathcal{F}\nu} \nonumber\\
&+ \frac{\mathcal{F}\alpha^4\nu y_2}{(1\!-\!\nu)c_4\!+\!\mathcal{F}\nu},
\label{eqn:Auxilliary}
\end{align}
where the initial conditions are estimated by substituting the free stream values of the unknown (i.~e., in the limit $y\rightarrow\infty$ substitute $\phi_1 \sim e^{-\alpha y}$ and $\phi_3 \sim e^{-Q y}$ in equation~\eqref{eqn:AuxiliaryVar}) and normalizing with respect to one of the variables (e.~g., $y_1$) to remove stiffness. The rescaled initial conditions for solving equations~\eqref{eqn:Auxilliary} are
\vspace{-0.5cm}
\begin{align}
& y_1 = 1.0, \quad y_2 = -(\alpha +q), \quad y_3 = \alpha^2 +q\alpha + q^2, \nonumber \\
& y_4 = q\alpha, \quad y_5 = -q\alpha(\alpha + q), \quad y_6 = (q\alpha)^2
\label{eqn:ICs}
\end{align}
The numerical solution for IVP~(\ref{eqn:Auxilliary}, \ref{eqn:ICs}) is obtained by marching backward from the free stream to the centerline. A suitable value of the eigenpair ($\alpha, \omega$) is obtained by enforcing the dispersion relation (equations~(\ref{eqn:OSE_DRP_AS}, \ref{eqn:OSE_DRP_S})) in auxiliary variables and solving simultaneously, i.~e.,
\bseq
\begin{align}
&\text{Re}(y_2) = 0 \quad \text{at} \quad y = 0 \quad \text{for odd component}, \label{eqn:DRP_AS} \\
&\text{Re}(y_5) = 0 \quad \text{at} \quad y = 0 \quad \text{for even component}, \label{eqn:DRP_S}
\end{align}
\label{eqn:DRPAux}
\eseq
respectively. $\text{Re}( \cdot )$ denotes the real part of the complex valued function. 

%%%%%%%%%%%%%%%%%%%%%%%%%%%%%%%%%%%%%%%%%%%%%
\section{\label{sec:level1}Results} \label{sec:results}
\noindent The zeros of the dispersion relation (equation~\eqref{eqn:DRPAux}) were explored within the complex $\alpha-\omega$ plane inside the region $-0.02 \le \omega_r \le 0.13, -1.6 \le \omega_i \le 0.25, \alpha_r \le 1.6$ and $|\alpha_i| \le 0.02$. Previous results indicate that the influence of viscoelasticity is fully captured by the modified elasticity number, $E = \frac{\nu We}{Re}$, a parameter representing the ratio of the fluid relaxation time to the characteristic time for vorticity diffusion~\cite{Ray2014}. We highlight our instability results versus this parameter. The continuation curves in \S \ref{subsec:tsa} and \ref{subsec:stsa} are depicted within the range $E \in [10^{-3}, \,\,5.0]$ using a discrete step-size of $\triangle E = 10^{-3}$. The numerical solution of the IVP~(\ref{eqn:Auxilliary}, \ref{eqn:ICs}) was determined via the fourth order Runge Kutta integration with a step-size of $\triangle y = 2.2 \times 10^{-3}$. The integration domain was truncated at $\eta = 12.0$ (a point at which the free stream boundary conditions~\eqref{eqn:FS} were imposed), which leads to a value of the momentum thickness (the reference length scale introduced in \S \ref{subsec:lsa}), $\delta = 0.30685$. The results in \S \ref{subsec:tsa} and \S \ref{subsec:stsa} are compared at two different values of $Re$ (i.~e., $Re=40$ and $Re=400$) as well as, at $\nu=0.3$ (the viscous stress dominated case) and at $\nu=0.7$ (the elastic stress dominated case).

%%%%%%%%%%%%%%%%%%%%%%%%%%%%%%%%%%%%%%%%%%%%%
\subsection{\label{sec:level2}Numerical method validation}\label{subsec:mv}
\begin{figure}[htbp]
\centering
\begin{subfigure}{0.48\textwidth}
\includegraphics[width=2\linewidth, height=2\linewidth, natwidth=610, natheight=642]{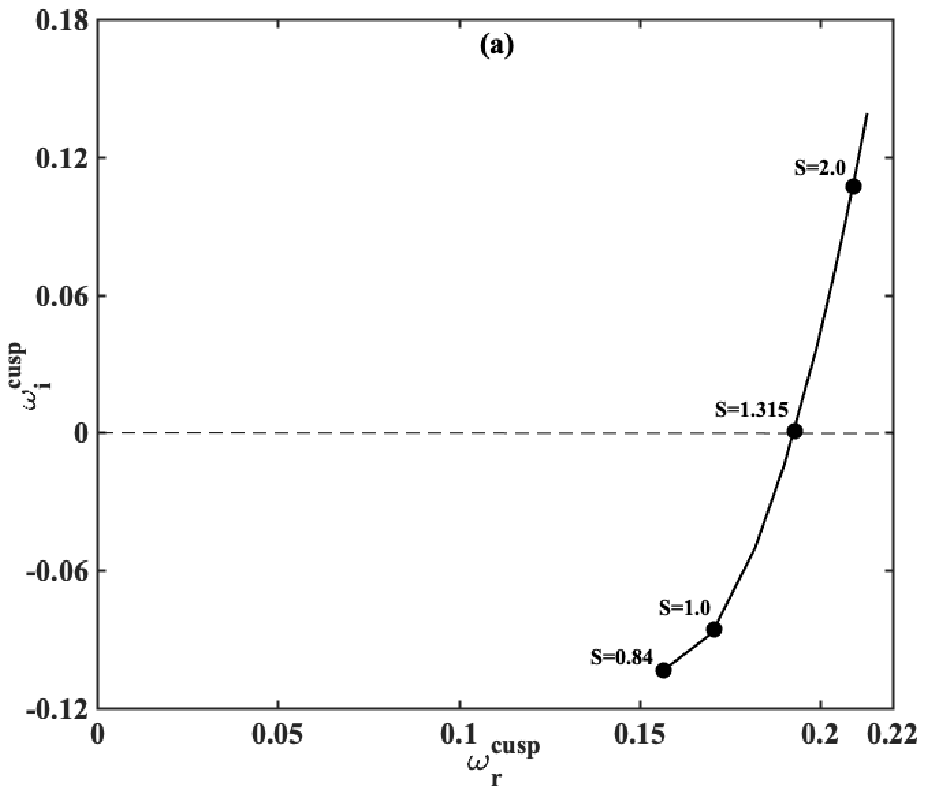}
\vskip -17pt
\caption*{} 
\end{subfigure}
\begin{subfigure}{0.48\textwidth}
 \includegraphics[width=2\linewidth, height=2\linewidth, natwidth=610, natheight=642]{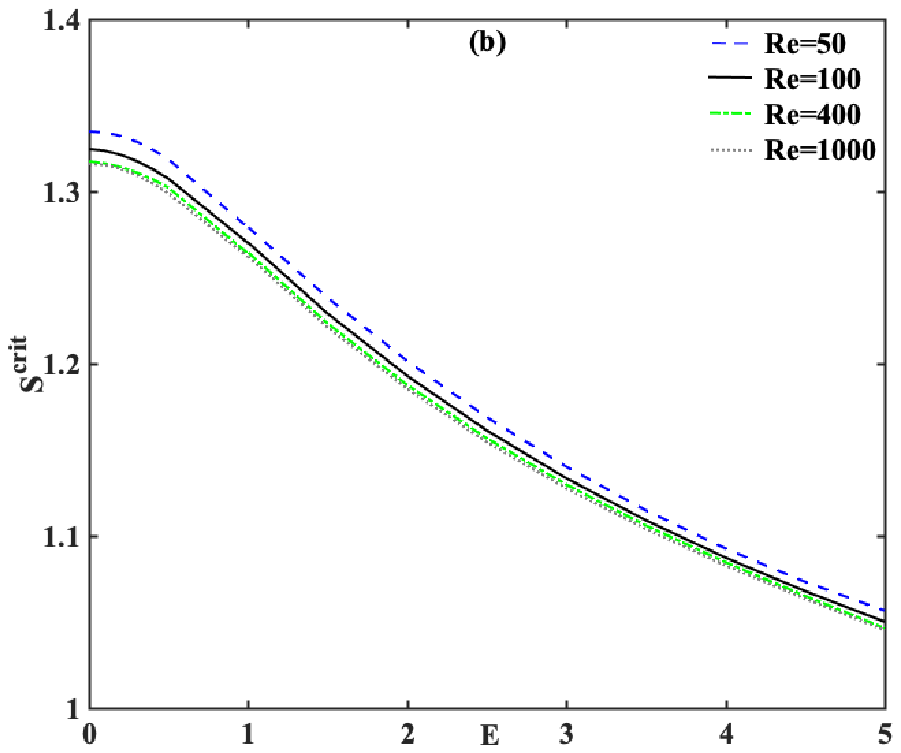}
\vskip -17pt
\caption*{} 
\end{subfigure}
\caption{(a) Locus of the cusp points of the Rayleigh equation in $\omega$-plane at critical values of the parameter, $S$, and (b) the critical values of $S$ versus elasticity number, $E$, of the OSE (equation~\eqref{eqn:OSE}) for Oldroyd-B fluids, at $\nu=0.5$ and $Re = 50, 100, 400, 1000$. The mean velocity profile is given by equation~\eqref{eqn:MeanflowHuerre}.}
\label{fig:Fig1}
\end{figure}

First, our numerical method outlined in \S \ref{sec:CMM} is validated by reproducing the absolute and the convective instability results for (a) inviscid mixing layers investigated by Huerre and Monkewitz~\cite{Huerre1985} or the Rayleigh instability equation (figure~\ref{fig:Fig1}a), and (b) spatially developing viscoelastic Oldroyd-B mixing layers probed by Ray and Zaki~\cite{Ray2014} (figure~\ref{fig:Fig1}b), using the non-dimensional base-state velocity profile given by
\beq
U(y) = 1 + S \tanh\left(\frac{y}{2}\right),
\label{eqn:MeanflowHuerre}
\eeq
where the parameter $S$ is the ratio of the difference and the sum of the free-stream velocities of the upper and the lower half. In figure~\ref{fig:Fig1}a, we recover the familiar curve of the cusp / pinch points for different complex pairs $(\alpha, \omega)$ determined by the numerical integration of the inviscid Rayleigh equation together with the exponentially decaying far stream boundary condition, equation~\eqref{eqn:FS}. Note the crossover from the real-$\omega$ axis at the critical value, $S=1.315$, highlighting a transition from convective instability (i.~e., $\omega_i^{\text{cusp}}<0$) to absolute instability at this value. Figure~\ref{fig:Fig1}b presents the evolution of the critical value of $S$ versus the elasticity number, $E$, at $\nu=0.5$ and $Re=50, 100, 400, 1000$ for an Oldroyd-B fluid. First, notice the reduction at low elasticity number regime, $E<0.1$, followed by an enhancement of absolute instability, with increasing $E$ (i.~e., the critical value of $S$ eventually drops with increasing $E$). Second, note that the elasticity number is  the dominant flow parameter measuring viscoelasticity while the Reynolds number has negligible influence beyond $Re \ge 400$. A more detailed outlook of the influence of viscoelasticity is acquired by examining the temporal growth rates, described in the next section. 
\vspace{-10cm}

%%%%%%%%%%%%%%%%%%%%%%%%%%%%%%%%%%%%%%%%%%%%%
\subsection{\label{sec:level2}Temporal stability analysis}\label{subsec:tsa}
\begin{figure}[htbp]
\centering
\begin{subfigure}{0.328\textwidth}
\includegraphics[width=2\linewidth, height=2\linewidth, natwidth=610, natheight=642]{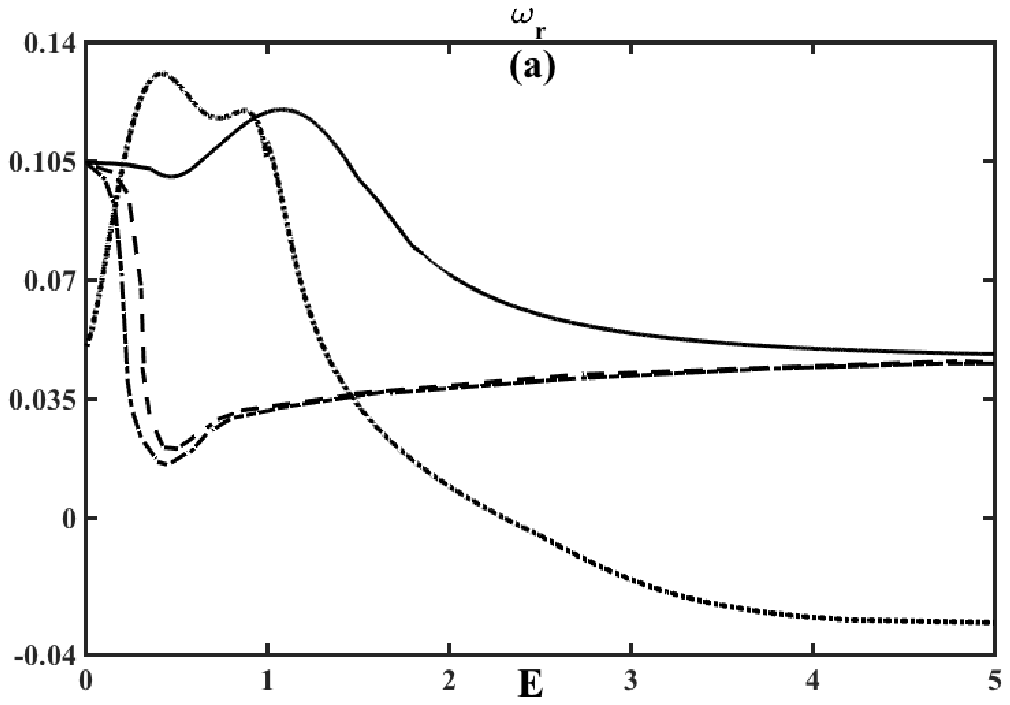}
\vskip -17pt
\caption*{} 
\end{subfigure}
\begin{subfigure}{0.328\textwidth}
\includegraphics[width=2\linewidth, height=2\linewidth, natwidth=610, natheight=642]{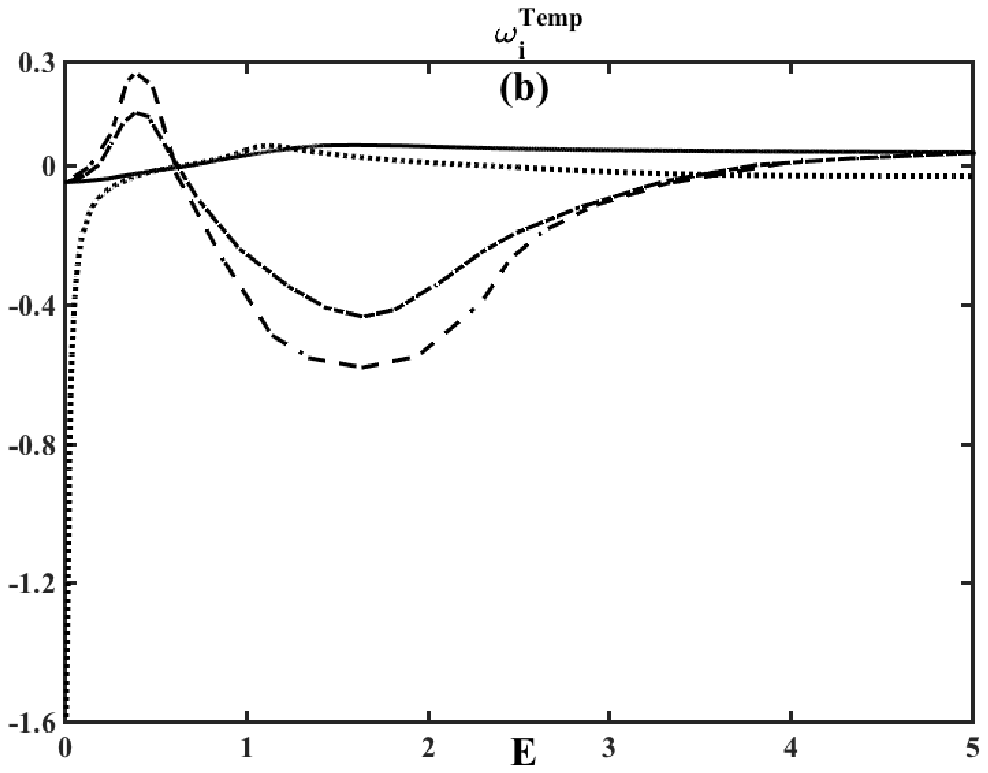}
\vskip -17pt
\caption*{} 
\end{subfigure}
\begin{subfigure}{0.328\textwidth}
\includegraphics[width=2\linewidth, height=2\linewidth, natwidth=610, natheight=642]{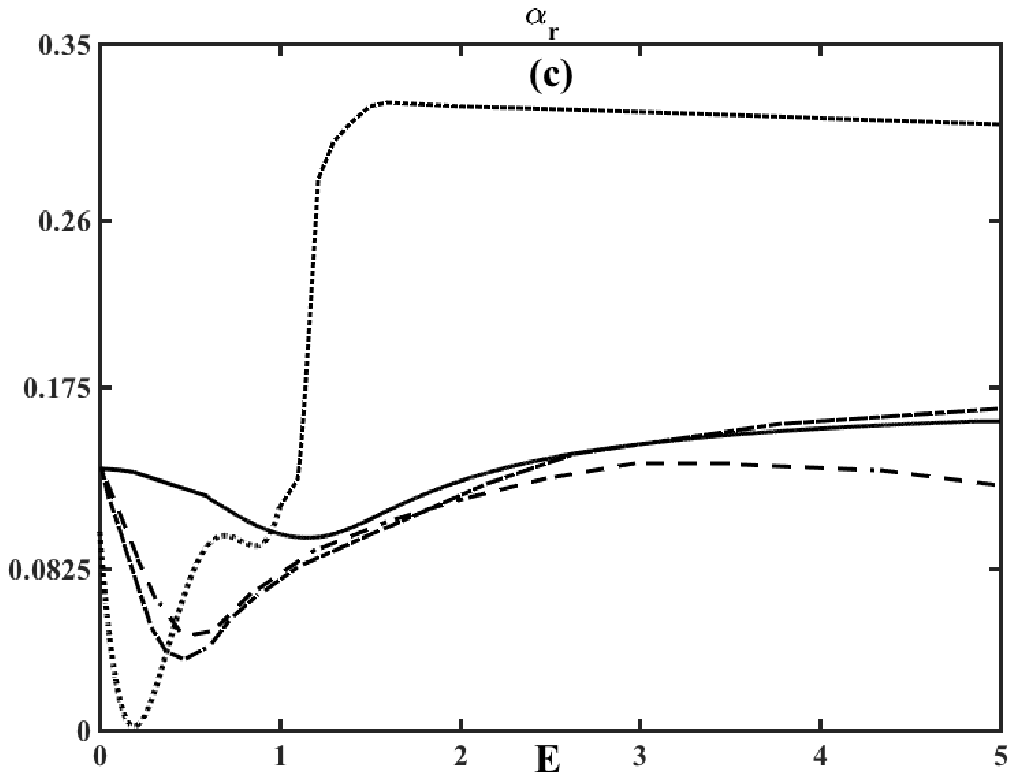}
\vskip -17pt
\caption*{} 
\end{subfigure}
%\vspace{-30cm}
\begin{subfigure}{0.328\textwidth}
\includegraphics[width=2\linewidth, height=2\linewidth, natwidth=610, natheight=642]{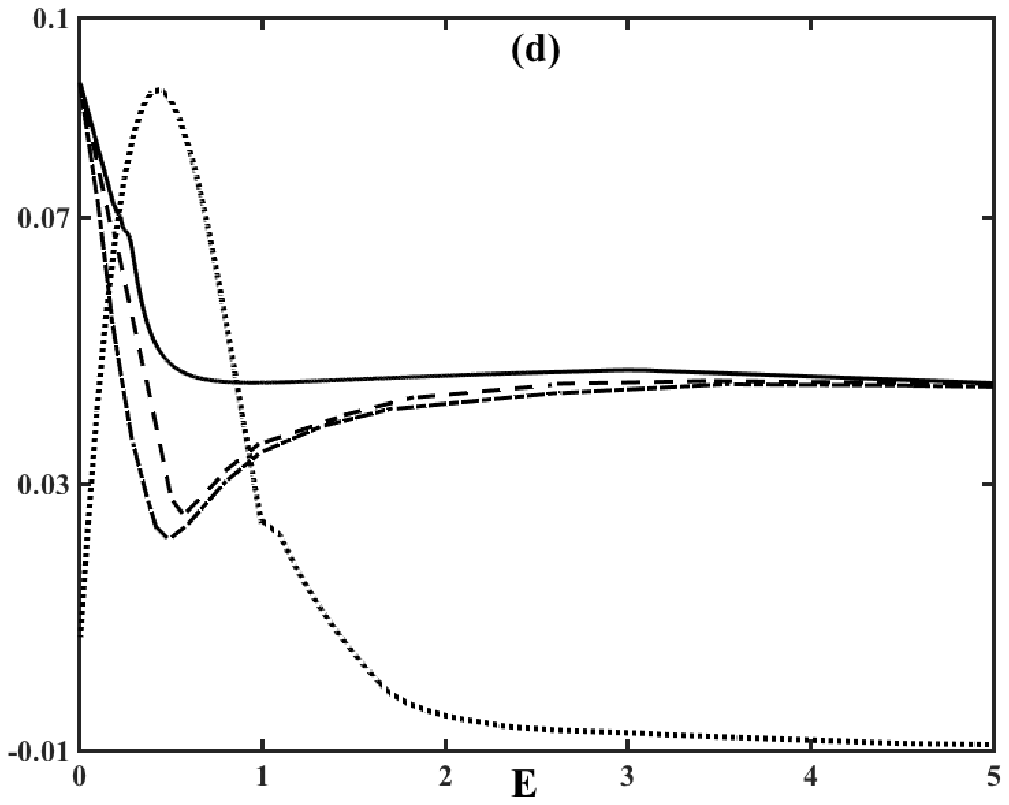}
\vskip -17pt
\caption*{} 
\end{subfigure}
\begin{subfigure}{0.328\textwidth}
\includegraphics[width=2\linewidth, height=2\linewidth, natwidth=610, natheight=642]{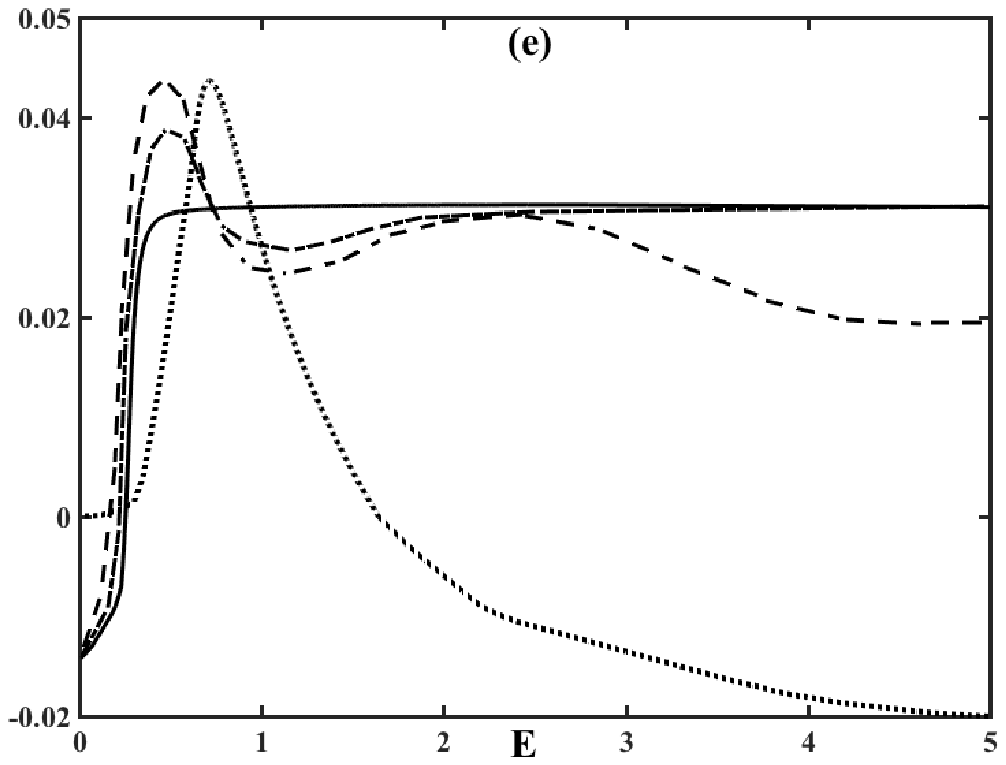}
\vskip -17pt
\caption*{} 
\end{subfigure}
\begin{subfigure}{0.328\textwidth}
\includegraphics[width=2\linewidth, height=2\linewidth, natwidth=610, natheight=642]{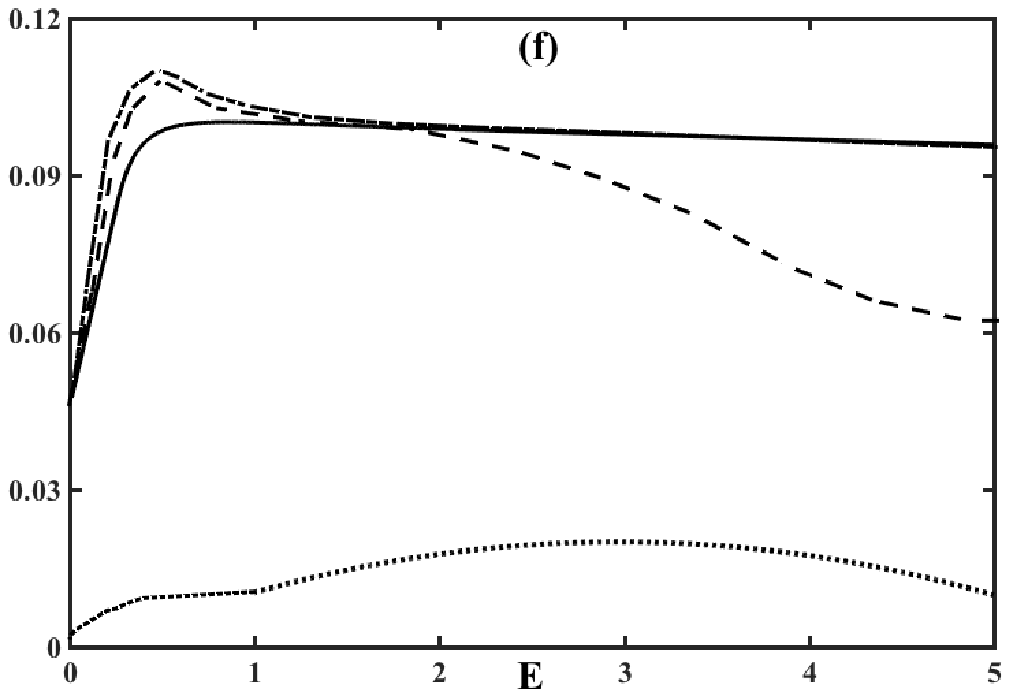}
\vskip -17pt
\caption*{} 
\end{subfigure}
\begin{subfigure}{0.328\textwidth}
\includegraphics[width=2\linewidth, height=2\linewidth, natwidth=610, natheight=642]{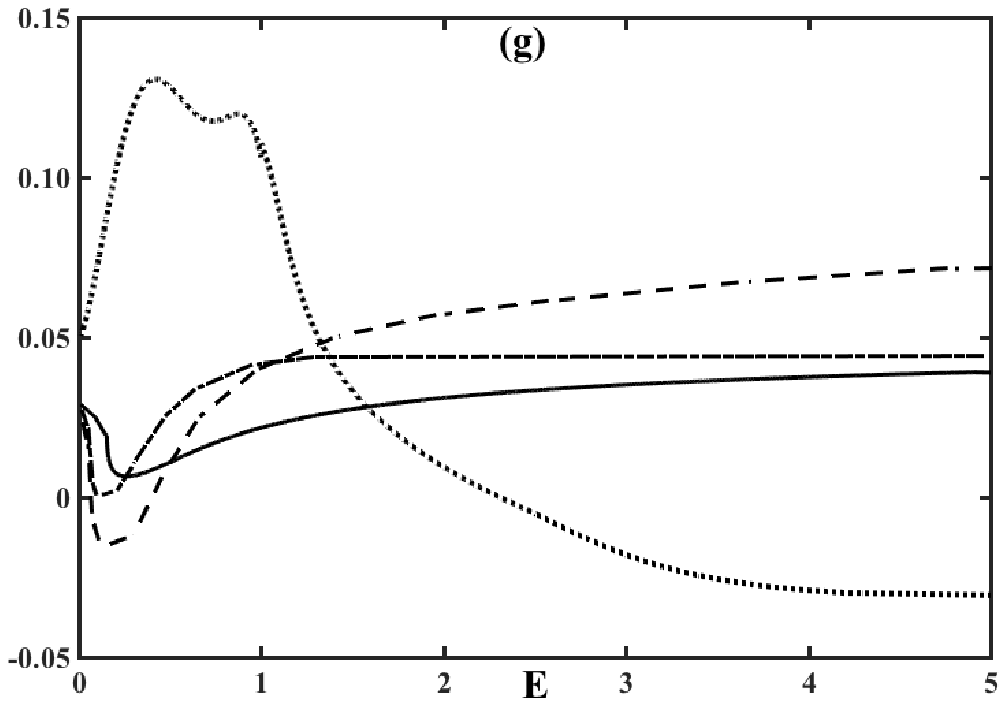}
\vskip -17pt
\caption*{} 
\end{subfigure}
\begin{subfigure}{0.328\textwidth}
\includegraphics[width=2\linewidth, height=2\linewidth, natwidth=610, natheight=642]{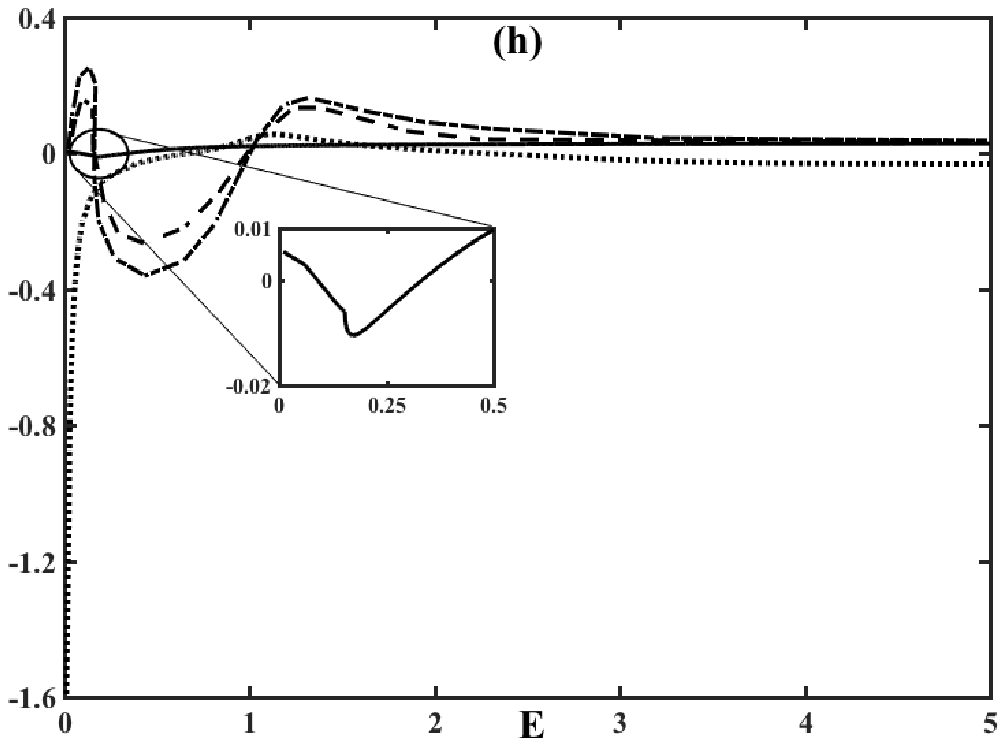}
\vskip -17pt
\caption*{} 
\end{subfigure}
\begin{subfigure}{0.328\textwidth}
\includegraphics[width=2\linewidth, height=2\linewidth, natwidth=610, natheight=642]{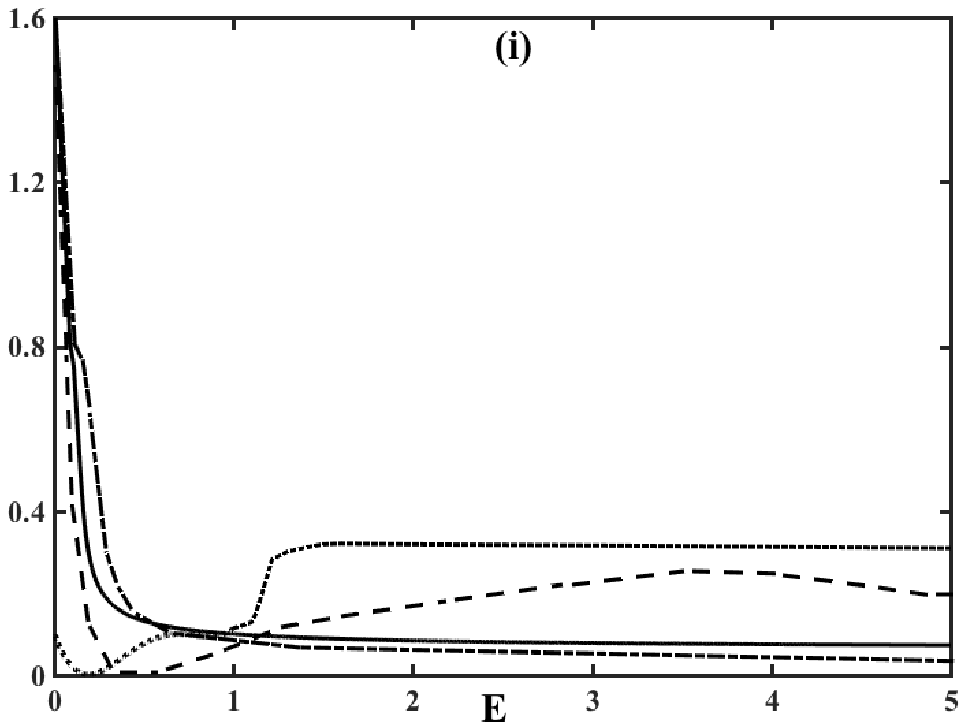}
\vskip -17pt
\caption*{} 
\end{subfigure}
\begin{subfigure}{0.328\textwidth}
\includegraphics[width=2\linewidth, height=2\linewidth, natwidth=610, natheight=642]{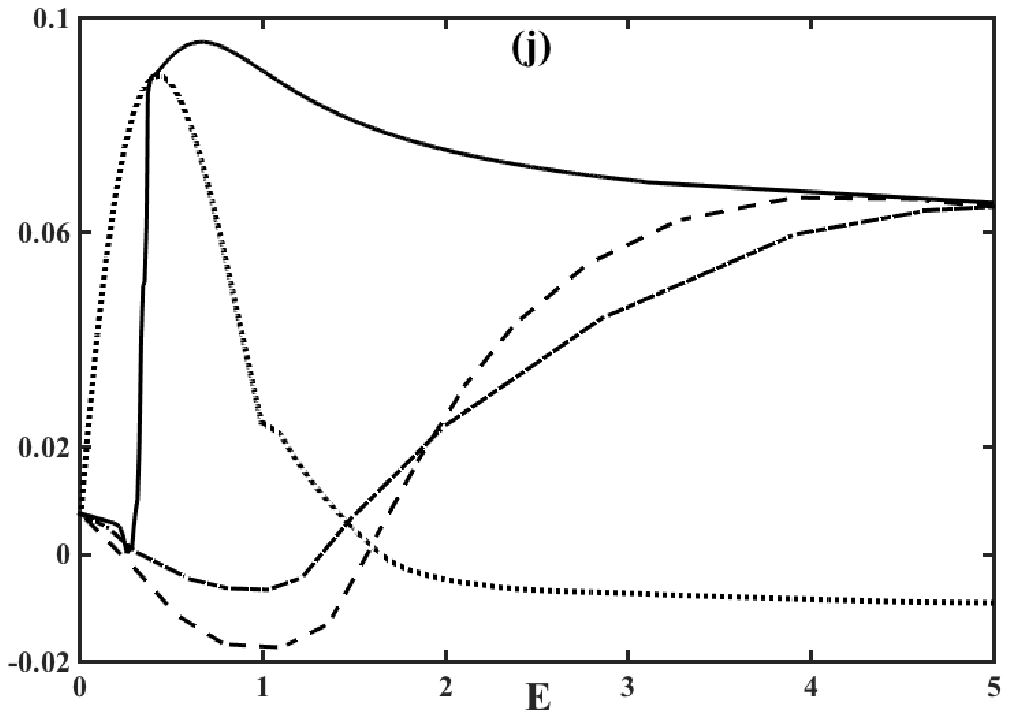}
\vskip -17pt
\caption*{} 
\end{subfigure}
\begin{subfigure}{0.328\textwidth}
\includegraphics[width=2\linewidth, height=2\linewidth, natwidth=610, natheight=642]{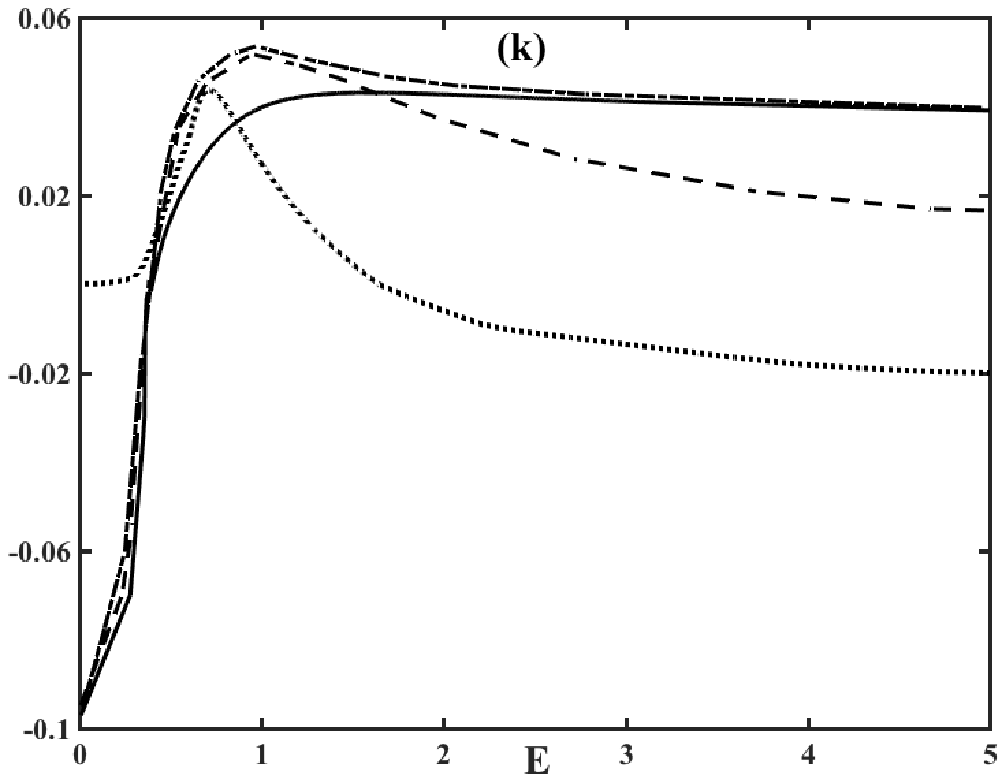}
\vskip -17pt
\caption*{} 
\end{subfigure}
\begin{subfigure}{0.328\textwidth}
\includegraphics[width=2\linewidth, height=2\linewidth, natwidth=610, natheight=642]{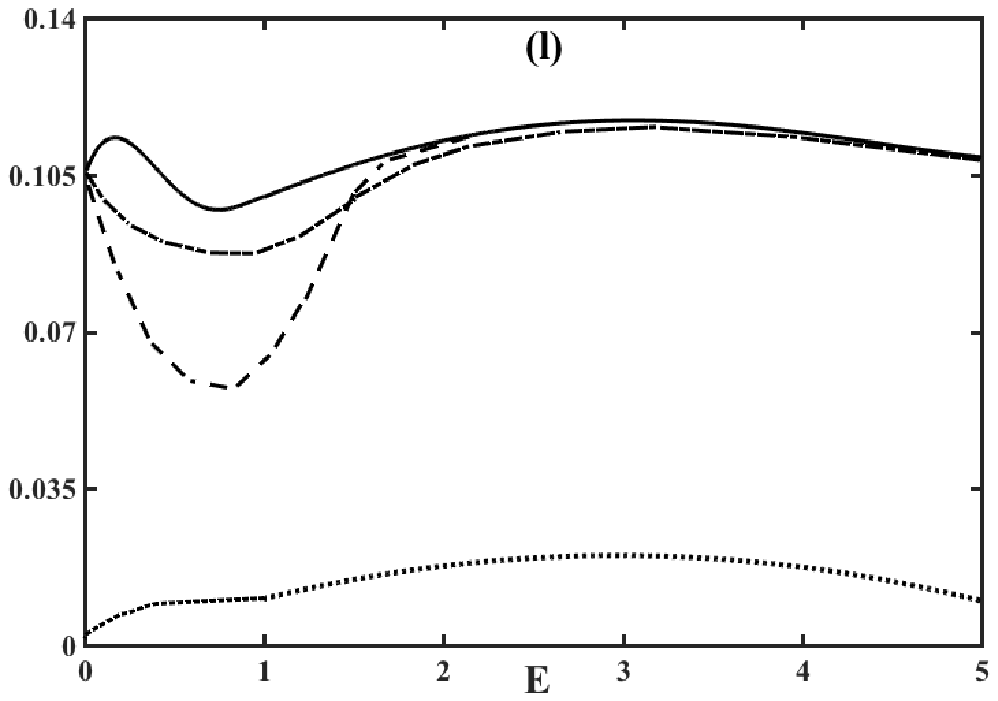}
\vskip -17pt
\caption*{} 
\end{subfigure}
\caption{The frequency, $\omega_r$, (first column); most unstable mode, $\omega^{\text{Temp}}_i$, (second column); wavenumber, $\alpha_r$ (third column) for parameters (a-c) $\nu=0.3$, $Re=40$, (d-f) $\nu=0.3$, $Re=400$, (g-i) $\nu=0.7$, $Re=40$ and (j-l) $\nu=0.7$, $Re=400$ versus the elasticity number, $E$, using using solid, dotted, dash-dot and dashed curves for Oldroyd-B, UCM, JS (at $a=0.5$) and PTT fluids (at $a=0.5, \varepsilon=0.5$), respectively.}
\label{fig:Fig2}
\end{figure}

A positive sign of the temporal growth rate indicates whether absolute instability is possible. The temporal stability analysis for viscoelastic free shear flows in the dilute regime, for low to moderate $Re$ and $We$, was earlier studied by us~\cite{Sircar2019}. 
In the limit of large $Re$ and $We$ such that $We/Re \sim \mathcal{O}(1)$, Azaiez conducted the temporal stability analysis through an elastic Rayleigh equation and concluded elasticity as the controlling flow parameter within the dilute flow regime~\cite{Azaiez1994}. In this study, we extend these ideas within the non-affine / non-monotonic response regime. Figure~\ref{fig:Fig2} presents the solution of the OSE~\eqref{eqn:OSE} for purely real wavenumbers, $\alpha_r$ (third column), while allowing angular frequency to be complex number. The other two columns in figure~\ref{fig:Fig2} are the temporal growth rate or the largest positive imaginary component of any root of the dispersion relation~\eqref{eqn:DRPAux}, $\omega_i^{\text{Temp}}$ (second column) and the corresponding real part of the frequency (first column). The four models are shown using solid line (Oldroyd-B fluids), dotted line (UCM fluids), dash-dot line (JS fluids, $a=0.5$) and dashed line (PTT fluids, $a=0.5, \varepsilon=0.5$).

First, notice the instability curves (in particular the curves for the UCM fluid) do not start at the same point near $E=0$. Readers are reminded that the simulation starts at $E_0=10^{-3}$, i.~e., $E \equiv 0$ (the Newtonian case) is not considered in the present discussion. Further note that the elastic stress dominated case at $Re=40$, with the exception of the UCM fluid, is unstable in the limit $E \rightarrow 0$ (figure~\ref{fig:Fig2}h). This instability occurs at short wavelength (i.~e., large $\alpha_r$, figure~\ref{fig:Fig2}i) and low frequency (i.~e., small $| \omega_r |$, figure~\ref{fig:Fig2}g). In the dilute regime, the shear flow instabilities were found to arise at zero elasticity number, through a combination of instability via normal stress anisotropy and elasticity~\cite{Sircar2019}. In the non-affine regime, we surmise a similar operative mechanism. The overall trend, for large values of $E$, is that elasticity is stabilizing for UCM fluids (figures~\ref{fig:Fig2}b,e,h,k), stabilizing for PTT fluids at higher $Re$ (figures~\ref{fig:Fig2}e,k) but has negligible small (although stabilizing) influence on Oldroyd-B and JS fluids ($\omega_i^{\text{Temp}}$ curve is nearly flat in the range $E \ge 3.5$). The mechanism of this elasticity-induced stabilization is well documented by Hinch in an appendix to Azaiez~\cite{Azaiez1994}, and is akin to the action of a `surface tension'. The stretched polymers contribute to an effective tension along the vibrating membrane that is the shear layer, and this tension damps the perturbations. This important analogy with the surface tension helps to provide a physical explanation for the influence of viscoelasticity on mixing layer stability, in the large-$E$ limit.

The JS as well as the PTT fluids at $Re=40$ shows a non-monotonic behavior at low to intermediate values of $E$ (figures~\ref{fig:Fig2}b,h) with alternating regimes of stability followed by instability, a feature typical of semi-dilute or moderately concentrated polymeric liquids. These fluctuations are characterized at short wavelengths (figures~\ref{fig:Fig2}c,i) and low frequency (figures~\ref{fig:Fig2}a,g). However, the PTT model `overestimates' (`underestimates') the instability predicted by the JS model for viscous stress dominated (elastic stress dominated) case at low to intermediate values of $E$ (i.~e., compare dashed and the dash-dot curves in figures~\ref{fig:Fig2}b,e versus figures~\ref{fig:Fig2}h,k). In addition, within the range $E \ge 3.5$, while the stability of the Oldroyd-B as well as the JS fluid is almost exclusively a function of $E$ (figures~\ref{fig:Fig2}b,e,h,k), the PTT model produces a more complicated behavior, dependent on all the three parameters, $E, Re, \nu$. The reason for these observations is explained as follows. The JS model is derived from a kinetic theory in which the polymers are represented as beads connected by springs and this underlying assumption of infinitely extensible springs is limiting in the JS model. The finitely extensible (nonlinear) springs of the PTT model arrests the infinite stresses at finite strain rate in extensional flows, leading to a complex pattern described above.
%\vspace{-20cm}

\begin{figure}[htbp]
\centering
\begin{subfigure}{0.48\textwidth}
\includegraphics[width=2\linewidth, height=2\linewidth, natwidth=610, natheight=642]{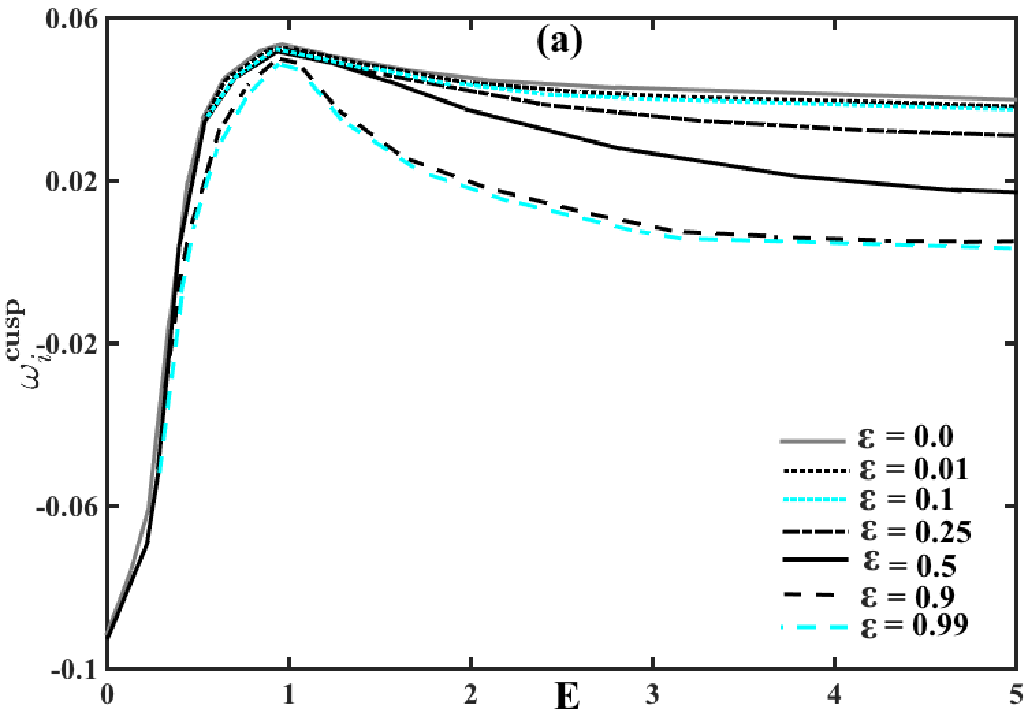}
\vskip -17pt
\caption*{} 
\end{subfigure}
\begin{subfigure}{0.48\textwidth}
\includegraphics[width=2\linewidth, height=2\linewidth, natwidth=610, natheight=642]{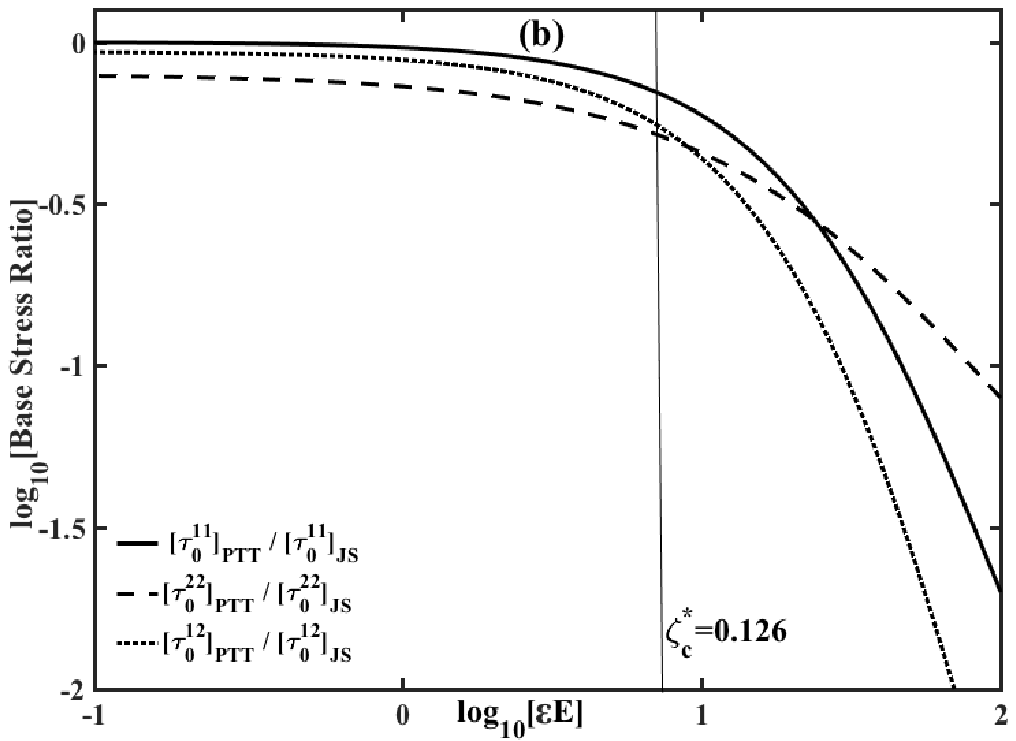}
\vskip -17pt
\caption*{} 
\end{subfigure}
\caption{(a) The temporal growth rate for linear PTT fluids, $\omega_i^{\text{Temp}}$ versus $E$, and (b) the log-log plot of the ratio of the PTT versus JS base stresses as a function of the centerline shear rate. Other parameters are fixed at $a=0.5, \nu=0.7, Re=400$ and $\varepsilon=0.5$ for plot (b).}
\label{fig:Fig3}
\end{figure}

In particular, the PTT results (compared with the JS results at $\nu=0.7, Re=400$, figure~\ref{fig:Fig3}a) indicate that a realistic value of the maximum elongation, $\varepsilon$, can induce a significant stabilizing influence of elasticity for large values of $E$. For small values of $\varepsilon$ (i.~e., $\varepsilon \le 0.1$), results are close to those obtained with the JS model. As $\varepsilon$ is increased, the (nonlinear) stiffness of the polymer molecules is increased, and one expects a reduction in the (stabilizing) influence of elasticity. We have considered several values of $\varepsilon$, but are particularly interested in results for $\varepsilon=0.5$. Previous analysis of viscoelastic mixing layers in high Weissenberg number flows have used values between $\varepsilon=0.3$ and $\varepsilon=0.7$~\cite{Hagen1997}. 

We gain a better understanding of these results and construct a criteria for the recovery of the `JS behavior' by examining the base-state polymer stresses, $\tau_0$ (equation~\eqref{eqn:baseStress}). In steady homogeneous shear flow, the behavior of the base-state polymer stress in the PTT mixing layer is closely related to the standard viscometric results which, in succession, depends on the non-dimensional shear rate~\cite{Ferras2019}. At low shear rates, the elongation of the PTT springs is modest, and the PTT and JS models produce similar results. At higher shear rates, the springs become more stretched, nonlinear elasticity becomes important, and the PTT polymer stress components become attenuated relative to their JS counterparts. These trends are present in the tanh mixing layer as well, though the velocity gradient is not constant, and the polymer stress depends on the local shear rate. The role of elongation must also be considered, and using Eqs.~\eqref{eqn:baseStress}, the ratio of the JS and PTT stress components can be shown to be functions of the shear rate, $\zeta \equiv \varepsilon E U_y$. Figure~\ref{fig:Fig3}b presents the base stresses at $\nu=0.7, Re=400$, with respect to this shear rate at the centerline, $\zeta = \zeta_c = \varepsilon E$. Unsurprisingly, when the shear rate, $\zeta_c$, is small, there is little difference between the PTT and the JS models, but at larger values of $\zeta_c$, the JS stress is larger. We can foresee that the relative attenuation of the base stresses as $\zeta_c$ is increased, is connected to the stabilizing influence of elasticity which was observed in our temporal stability results. Conversely, we would expect to recover the JS behavior if $\zeta_c$ is less than some critical value, $\zeta^*_c=0.126$, i.~e., the ratio of all the base stresses are less than 90\% beyond this critical value of $\zeta_c$. A better characterization of these instabilities are revealed through the spatiotemporal analysis in \S \ref{subsec:stsa}.
\vspace{-10cm}

%%%%%%%%%%%%%%%%%%%%%%%%%%%%%%%%%%%%%%%%%%%
\subsection{\label{sec:level2}Spatiotemporal stability analysis}\label{subsec:stsa}
\begin{figure}[htbp]
\centering
\begin{subfigure}{0.48\textwidth}
\includegraphics[width=2\linewidth, height=2\linewidth, natwidth=610, natheight=642]{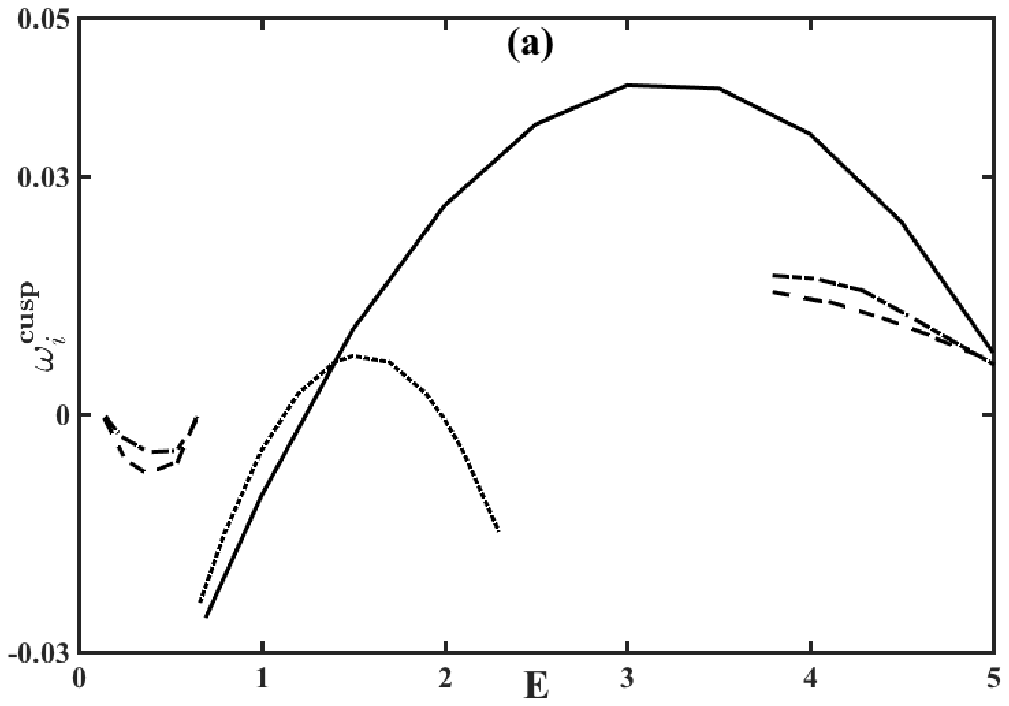}
\vskip -17pt
\caption*{} 
\end{subfigure}
\begin{subfigure}{0.48\textwidth}
\includegraphics[width=2\linewidth, height=2\linewidth, natwidth=610, natheight=642]{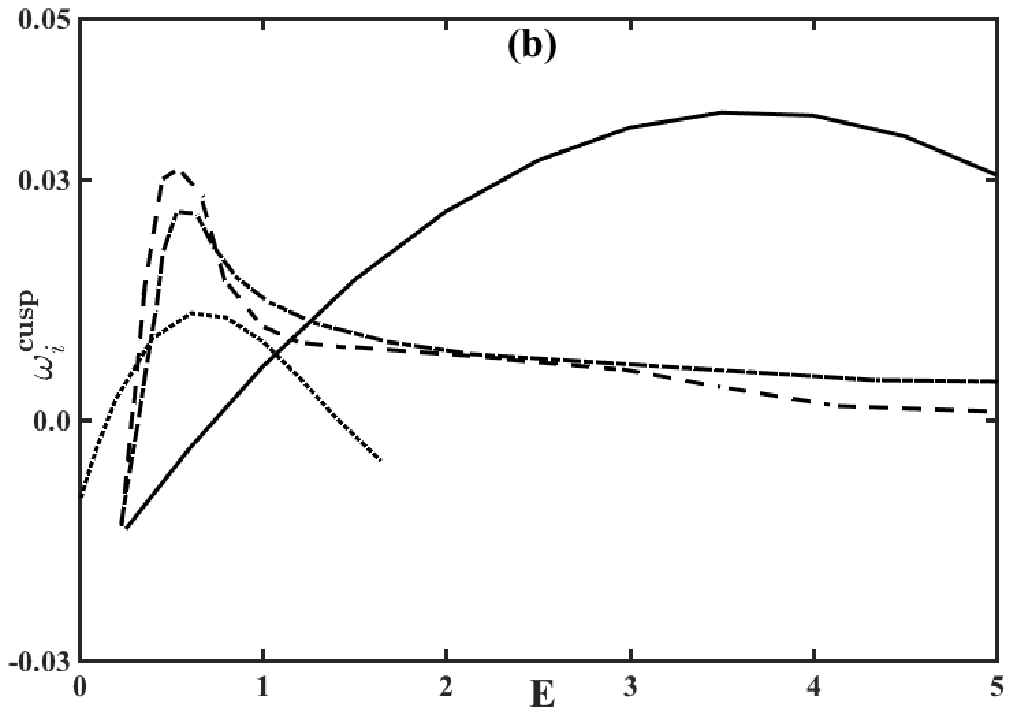}
\vskip -17pt
\caption*{} 
\end{subfigure}
%\vspace{-5cm}
\begin{subfigure}{0.48\textwidth}
\includegraphics[width=2\linewidth, height=2\linewidth, natwidth=610, natheight=642]{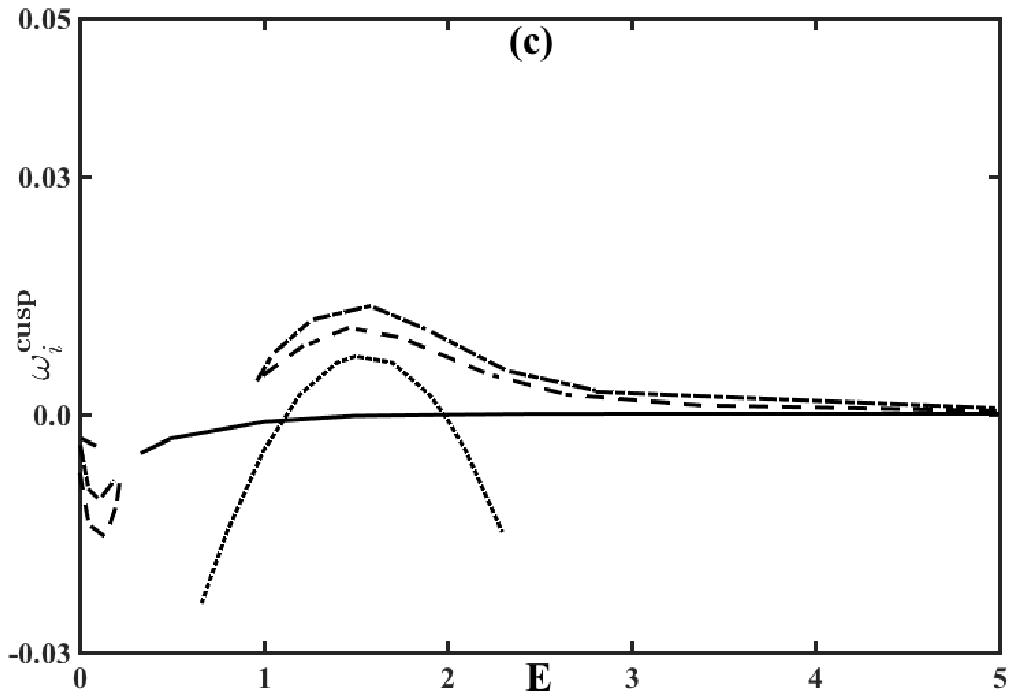}
\vskip -17pt
\caption*{} 
\end{subfigure}
\begin{subfigure}{0.48\textwidth}
\includegraphics[width=2\linewidth, height=2\linewidth, natwidth=610, natheight=642]{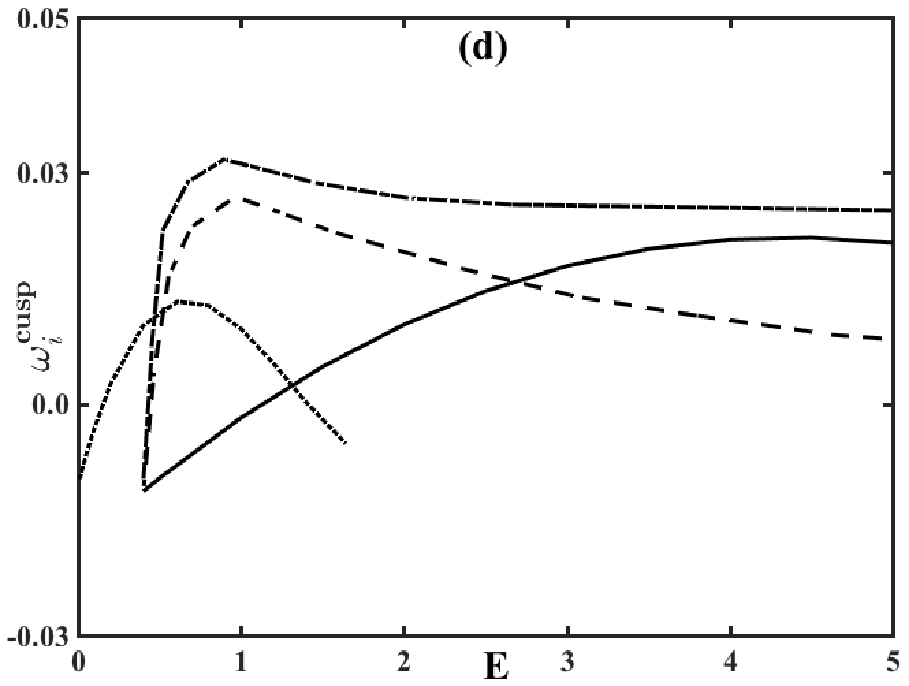}
\vskip -17pt
\caption*{} 
\end{subfigure}
\caption{The cusp point, $\omega^{\text{cusp}}_i$ versus $E$ evaluated at the flow parameters, (a) $\nu=0.3, Re=40$, (b) $\nu=0.3, Re=400$, (c) $\nu=0.7, Re=40$, and (d) $\nu=0.7, Re=400$, shown using solid, dotted, dash-dot and dashed curves for Oldroyd-B, UCM, JS (at $a=0.5$) and PTT fluids (at $a=0.5, \varepsilon=0.5$), respectively.}
\label{fig:Fig4}
\end{figure}

Spatiotemporal analysis is typically relevant when one introduces an impulse excitation locally in a flow and observes how that disturbance evolves. In an effort to determine the range of $E$ (for fixed $Re$ and $\nu$) for which the flow regimes are absolutely unstable, convectively unstable or temporally stable, we recover the absolute growth rate (or the growth rate at the cusp point, $\omega^{\text{cusp}}_i$, figure~\ref{fig:Fig4}) starting from the most unstable temporal mode, $\omega^{\text{Temp}}_i$. The cusp point in the $\omega-$plane is a saddle point satisfying the criteria, $D(\alpha, \omega^{\text cusp})\!=\!\frac{\partial D(\alpha, \omega^{\text cusp})}{\partial \alpha}\!=\!0$ but $\frac{\partial^2 D(\alpha^{\text pinch}, \omega^{\text cusp})}{\partial \alpha^2}\!\ne \!0$ (where $D(\alpha, \omega)=0$ is the dispersion relation). However, not all cusp points are unstable and, in particular, the evanescent modes are segregated from the regular cusp points using the Briggs idea of analytic continuation~\cite{Kupfer1987}. While the Oldroyd-B model is presumed to represent the dilute polymeric liquids, the linear PTT model (due to the finite, nonlinear elongation of the polymer chains at $a=0.5, \varepsilon=0.5$) portrays the instability transition for moderately concentrated polymeric liquids. A discontinuity in the curves in figure~\ref{fig:Fig4} indicates a region of temporal stability.

The instability pattern shown in figure~\ref{fig:Fig4} is a result of a complex interplay between the inertial forces (proportional to $Re$) and the normal stress anisotropy through elasticity (proportional to $E$). For example, within the lower elastic number regime (i.~.e., $E<0.5$) the dilute polymeric liquids display constricted regions of temporal stability at higher value of $Re$ (comparing the solid curve at $Re=40$ (figure~\ref{fig:Fig4}a) versus the curve at $Re=400$ (figure~\ref{fig:Fig4}b)). Similarly, the elastic stress dominated case reveals (convective) instability at lower value of $Re$ and $E$ (figure~\ref{fig:Fig4}c). Clearly, while the former observation is the result of inertia, the latter is the outcome of instability generated via polymer elasticity. Analogous with the temporal stability analysis, we find that the absolute growth rate of the PTT fluids in the high elastic number regime is always lower than the JS fluids. This is because the finite elongation attribute of the PTT fluids is able to rein in the large elastic stress gradient buildup.

\begin{figure}[htbp]
\centering
\begin{subfigure}{0.48\textwidth}
\includegraphics[width=2\linewidth, height=2\linewidth, natwidth=610, natheight=642]{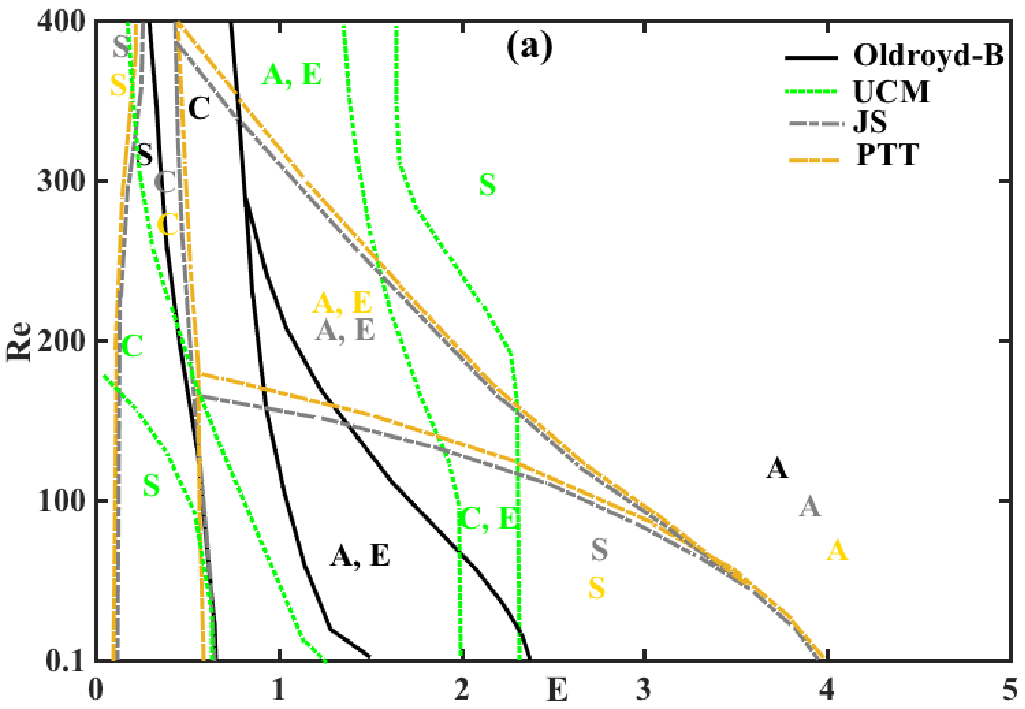}
\vskip -17pt
\caption*{} 
\end{subfigure}
\begin{subfigure}{0.48\textwidth}
\includegraphics[width=2\linewidth, height=2\linewidth, natwidth=610, natheight=642]{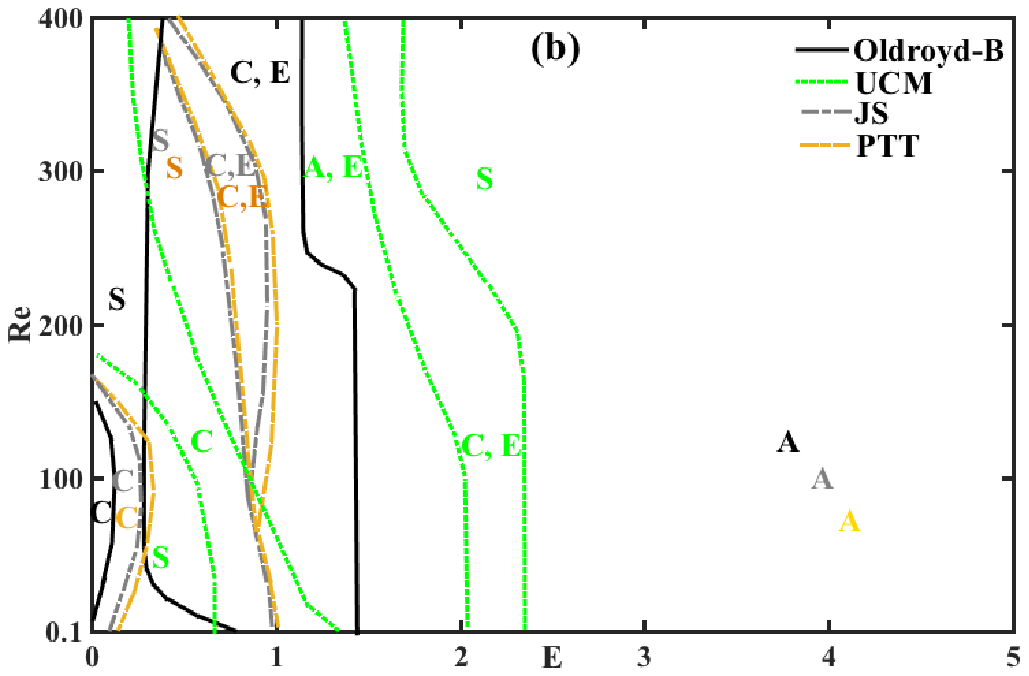}
\vskip -17pt
\caption*{} 
\end{subfigure}
\caption{Viscoelastic free shear layer stability phase diagram at (a) $\nu=0.3$, and (b) $\nu=0.7$, in the $E$--$Re$ parametric space. The regions {\bf S}, {\bf C} and {\bf A} are denoted by temporally stable, convectively unstable and absolutely unstable regions, respectively. The domains outline by ({\bf A}, {\bf E}) and ({\bf C}, {\bf E}) are those where both the stable and evanescent modes (denoted by {\bf E}) are found.}
\label{fig:Fig5}
\end{figure}

To explore the nature of these instabilities, we compute the boundaries of the temporally stable regions ({\bf S}), convective instabilities ({\bf C}), evanescent modes ({\bf E}) and absolute instabilities ({\bf A}) within a selected range of flow-elasticity parameter space, i.~e., $Re \in [0.1, 400], E \in [10^{-3}, 5]$ and $\nu = 0.3$ (figure~\ref{fig:Fig5}a) and $\nu=0.7$ (figure~\ref{fig:Fig5}b). The boundaries of the temporally stable, convectively unstable and absolutely unstable region, for dilute liquids in the viscous stress dominated case (figure~\ref{fig:Fig5}a) are numerically estimated to reside within the range $E<0.3$, approximately $0.3 \le E \le 1.5$ and $E > 1.5$ (for all values of $Re$), respectively. In contrast, the concentrated polymeric liquids exhibit temporal stability and absolute instability in a confined region, i.~e., $E<0.1$ and $E>4$, respectively. For the elastic stress dominated case (figure~\ref{fig:Fig5}b), the dilute as well as the concentrated liquid reveals convective instability within the range $Re<150$ and $E<0.1$ (including in the limit $Re \ll 1$) which is followed with alternating regions of (temporal) stability and (convective) instability for larger values of $E$. Absolute instability is unveiled by the Oldroyd-B fluids for $E>1.5$ and by the PTT fluids for $E>1.0$, for practically all values of $Re$.
%\vspace{-50cm}

We discuss the significance of the temporal stability analysis (figure~\ref{fig:Fig2}) as well as the spatiotemporal phase diagram (figure~\ref{fig:Fig5}) in relation to the experiments for Newtonian as well as viscoelastic flow past a cylinder, since the shear layer instability is closely related to the instability of the cylinder wakes~\cite{Pipe2005}. In the case of Newtonian wakes for $Re<1$, the flow past a cylinder is steady and without recirculation~\cite{Taneda1956}. The emergence of a locally, convectively unstable wake is detected at $Re \approx 5$, whereby the selected perturbation are amplified and convected downstream but ultimately leave the flow undisturbed~\cite{Monkewitz1988}. The locally most unstable part of the wake becomes absolutely unstable at $Re \approx 25$, although this is not sufficient for self-sustained global oscillations of the wake. At $Re \approx 47$ the region of absolute instability in the wake is large enough for the wake to sustain time-amplified oscillations, which is followed by the onset of the laminar two-dimensional von K\'{a}rm\'{a}n instability~\cite{Chomaz1988}. This transition to global instability has been ascribed to a supercritical Hopf bifurcation towards a limit cycle~\cite{{Provansal1987}} and the linear stability analysis appears to faithfully describe the Newtonian wake dynamics, `unreasonably' far above the global instability threshold of $Re$~\cite{Oertel1990}.

In contrast, first notice that all four stress constitutive relations in the phase diagram in figure~\ref{fig:Fig5} indicates a convectively unstable region for moderate values of $E$. For the viscous stress dominated case, figure~\ref{fig:Fig5}a, this region lies in the range $0.6 \le E \le 0.7$ (Oldroyd-B), $E_0 \le E \le 0.1$ (UCM), $0.5 \le E \le 0.6$ (JS and PTT); and for the elastic stress dominated case, figure~\ref{fig:Fig5}b, this region lies in the range $0.8 \le E \le 1.1$ (Oldroyd-B), $E_0 \le E \le 0.1$ (UCM), $0.5 \le E \le 0.6$ (JS and PTT); for significantly larger values of $Re$ (i.~e., $Re>47$). This observation is in congruence with the early experimental studies of viscoelastic vortex street highlighting the reduction in the vortex shedding frequency (related to the temporal growth rate in the present study)~\cite{Kalashnikov1970}, as well as in the intensity of the vorticity~\cite{Cadot2000}, namely the origin of inertial turbulence modified by elasticity. Second, notice the appearance of convectively unstable region for the elastic stress dominated case, for small values of $E$ and in the limit $Re \rightarrow 0$. This result corroborates the experimental findings of McKinley \emph{et al.}~\cite{McKinley1993} showing instability induced by elasticity for viscoelastic flow past a cylinder at $Re \ll 1$ as well as the findings of Coelho and Pinho~\cite{Coelho2003} showing a significant destabilization of the wake for shear-thinning, elastic fluids at $Re<40$, in other words the presence of elastic turbulence.

Two other observations are noteworthy: first, the Oldroyd-B, JS and the PTT models suggest absolute instability for sufficiently large values of $E$ (i.~e., $E>4$ ($E>1.5$) for the viscous (elastic) stress dominated case, figure~\ref{fig:Fig5}a versus \ref{fig:Fig5}b) and second, the presence of a temporally stable region for the JS and PTT models for the viscous stress dominated case (figure~\ref{fig:Fig5}a), approximately within the range $0.6 < E < 4$ and $Re < 100$. While the former observation signifies a pathway to elastoinertial turbulence (appearing at moderate $Re$ and large $E$) which characterizes the maximum drag reduction state and even originates in linear instability studies of pipe flows~\cite{Chaudhary2021}, the latter observation indicates an intricate tug-of-war between the inertial destabilization and elastic stabilization. While the presence of the first region was recently substantiated~\cite{Samanta2013}, a comprehensive experimental ratification of the second region is eagerly awaited.

%\vspace{-1.5cm}
%%%%%%%%%%%%%%%%%%%%%%%%%%%%%%%%%%%%%%%%%%%%%
\section{Conclusions} \label{sec:conclusion}
%\vspace{-0.5cm}
This investigation addresses the linear, temporal and the spatiotemporal analyses of free shear flows of dilute as well as moderately concentrated polymeric liquids for low to moderate Reynolds number and Elasticity number, by capturing the non-affine flow response of a `tanh' base flow mixing velocity profile. Section~\ref{sec:mm_lsa} presented the viscoelastic free shear flow model as well as the elements of the linear stability analysis via the solution of the Orr-Sommerfeld equation. Section~\ref{sec:CMM} demonstrated the steps of the Compound Matrix Method, utilized to numerically solve the resultant system of stiff differential equations.
The temporal stability analysis in section~\ref{subsec:tsa}, indicates (a) elastic stabilization at higher values of elasticity number and (b) a non-monotonic instability pattern at low to intermediate values of elasticity number for the JS as well as the PTT model. The spatiotemporal phase diagram in section~\ref{subsec:stsa} divulge the familiar regions of inertial and elastic turbulence, a recently verified region of elastoinertial turbulence and the unfamiliar temporally stable region for intermediate values of Reynolds and Elasticity number.

Although this study provides an improved understanding of the linear dynamics of mixing layers for dilute to moderately concentrated polymeric liquids, a number of simplifying assumptions were made, and the relaxation of these assumptions paves a way for further progress. Understanding the importance of base flow spreading, confinement and nonlinearity as well as the consideration of the shear flows of polymer melts (or fluids with very large viscosity) are of substantial importance. Finally, we note that we have considered only the two-dimensional instabilities which is a common simplification in the absolute / convective instability studies of Newtonian flows where Squire's transformation can be applied~\cite{Squires2005}. A modified Squire's transformation for Oldroyd-B fluids also exist~\cite{Bistagnino2007}, but we are not aware of a similar result for the PTT model. Hence, a consideration of the three-dimensional instability modes in future studies with the PTT model may be worthwhile.
\vskip 5pt
\noindent \textbf{Acknowledgments} D. B. and S.S. acknowledges the financial support of the Grant CSIR 09 / 1117 (0004) / 2017-EMR-I and DST ECR / 2017 /000632, respectively.
%\begin{acknowledgments}
%\end{acknowledgments}

%\vspace{-0.8cm}
%%%%%%%%%%%%%%%%%%%%%%%%%%%%%%%%%%%%%%%%%%%%%
\appendix
\section{coefficients of OSE}  \label{sec:appendix}
\noindent The OSE, outlined in \S \ref{subsec:lsa}, coupled with the linear PTT stress constitutive equation, is given by
\beq
\left\{i \big[(\alpha U-\omega)(D^2 - \alpha^2) - \alpha D^2U \big] - \frac{1-\nu}{Re} (D^2 - \alpha^2)^2 \right\} \phi = \frac{\nu}{\mathcal{F} Re} \sum_{n=0}^{4} c_n D^n \phi,
\eeq
where the coefficients $c_i$'s are specified as follows
\bseq
\begin{align}
c_0 &= \frac{-4\left(S_3+(1+a)S_1\right)^2}{(4S_0^2d_0^3(S_3+(1+a)S_1)^2)} \left[ \left[S_0^2\left\{(D^2\tilde{A})d_0^2-\tilde{A}d_0(D^2d_0)-2(D\tilde{A})d_0(Dd_0) + 2\tilde{A}(Dd_0)^2+\alpha^2\tilde{A}d_0^2\right\} \right. \right. 
\nonumber \\
& + 2(1-a^2)i\alpha \left\{S_0\left((D^2U)(S_1+S_3)+(DU)(DS_1 + DS_3)\right)- (DU)(S_1+S_3)(DS_0) \right\}(\tilde{A})d_0^2+ 2(1- \nonumber \\
&\left. a^2)i\alpha(DU)(S_1+S_3)S_0\left\{(D\tilde{A})d_0^2-2\tilde{A}d_0(Dd_0)\right\}\right] + d_0^3\left[2S_0(S_3 +(1+a)S_1)\left\{2(S_1+S_3)(S_2-(1+ \right. \right. \nonumber\\
& \left. a)S_1)+1+2(S_3+(1+a)S_1)(S_1+S_3)\right\}i\alpha(1+a)(D^2\tau_0^{22})+ \left\{(2(S_1+S_3)(S_2-(1+a)S_1) + 1) - \right. \nonumber \\
&\left. 2(S_1+S_3)(S_3 + (1+a)S_1) \right\} i\alpha(1-a)(D^2\tau_0^{11})+\left\{(2(S_1 + S_3)(S_2 - (1+a)S_1) + 1)a + (S_1+S_3) \right. \nonumber \\
&\left. (S_3+(1+a)S_1)(1+a^2)\right\}4\alpha^2(D\tau_0^{12})+2\left\{2S_0(S_3+(1+a)S_1)((DS_1+DS_3)(S_2-(1+a)S_1)+ \right. \nonumber \\
& (S_1+S_3)(DS_2-(1+a)DS_1)) -(2(S_1+S_3)(S_2-(1+a)S_1)+1) ((DS_0)(S_3+(1+a)S_1)+ S_0 \nonumber\\
& \left. (DS_3 + (1+a)DS_1)) \right\}(i\alpha((1+a)D\tau_0^{22} + (1-a)D\tau_0^{11})+4a\alpha^2\tau_0^{12})+ 4\left\{S_0(S_3+(1+a)S_1)^2\right.  \nonumber \\
& \left. (DS_1+DS_3)-(S_3+(1+a)S_1)(S_1+S_3) ((DS_0)(S_3+(1+a)S_1) + 2S_0(DS_3 + (1+a)DS_1))\right\} \nonumber \\
& (i\alpha((1+a)(D\tau_0^{22}) - (1-a)(D\tau_0^{11})) + 2(1+a^2)\alpha^2\tau_0^{12})+4S_0(S_3 +(1+a)S_1)(S_1+S_3)(DS_3 \nonumber\\
& \left. \left.+ (1+a)DS_1) (i\alpha((1+a)D\tau_0^{22}-(1-a)D\tau_0^{11})+2a\alpha^2\tau_0^{12} )\right] \right],
\label{eqn:c0} 
\end{align}
\begin{align}
c_1 &= \frac{-4( S_3 +(1+a)S_1 )^2}{(4S_0^2d_0^3(S_3 +(1+a)S_1)^2)} \left[ \left[ (S_0)^2 \left\{ 2(D\tilde{A})d_0^2 + (D^2\tilde{B})d_0^2 - \tilde{B}d_0(D^2d_0) - 2(\tilde{A})d_0(Dd_0) - 2(D\tilde{B}) \right. \right. \right. \nonumber\\
& \left. d_0(Dd_0) + 2(\tilde{B})(Dd_0) + \alpha^2\tilde{B}d_0^2 \right\}+ 2(1-a^2)(i\alpha)\left\{S_0((D^2U)(S_1+S_3) + (DU)(DS_1 + DS_3))-  \right. \nonumber\\
& \left. \left. (DU)(S_1+S_3)(D(S_0)\right\}\tilde{B}d_0^2 + 2(1-a^2)(i\alpha)(DU)(S_1+S_3)S_0\left\{(\tilde{A} + D\tilde{B})d_0^2-(\tilde{B})d_0(Dd_0)\right\} \right] +  d_0^3  \nonumber\\
& \left[ 2S_0(S_3 +(1+a)S_1) \left\{(2(S_1+S_3)(S_2 - (1+a)S_1) + 1 + 2(S_3 + (1+a)S_1)(S_1+S_3) )(i\alpha)(1+a)(1 \right. \right. \nonumber\\
& - 2a) D\tau_0^{22}+ (2(S_1+S_3)(S_2 - (1+a)S_1) + 1 - 2(S_1+S_3)(S_3 + (1+a)S_1) )i\alpha(1-a)(1+2a)D\tau_0^{11}  \nonumber\\
& \left. + (2a(S_1 + S_3)(S_2 - (1+a)S_1) + (S_1+S_3)(S_3+(1+a)S_1)(1+a^2))4\alpha^2\tau_0^{12}\right\} + 2\left\{(DS_1+DS_3) \right. \nonumber\\
& (S_2-(1+a) S_1) +(S_1 + S_3)(DS_2 - (1+a)DS_1) - ((DS_0)(S_3+(1+a)S_1) + S_0(DS_3 + (1+a)DS_1) \nonumber\\
&\left. ) \right\}\!\! (2(S_1\!\!+\!\!S_3)(S_2\!\!-\!\! (1+a)S_1) \!\!+\!\! 1)(2ai\alpha((1-a)\tau_0^{11} - (1+a)\tau_0^{22}) - \frac{4ai\alpha \nu}{ReE}) -4\left\{((DS_1+DS_3)(S_3 \right. \nonumber\\
& +(1+a)S_1) + (S_1+S_3)(DS_3+(1+a)DS_1)) - ((DS_0(S_3+(1+a)S_1) + S_0(DS_3 + (1+a)DS_1))(S_1 \nonumber\\
& \left. \left. +S_3)(S_3+(1+a)S_1)\right\}(2ai\alpha((1-a)\tau_0^{11} + (1+a)\tau_0^{22}) + \frac{4i\alpha \nu}{ReE} ) \right],
\label{eqn:c1} 
\end{align}
\begin{align}
c_2 &= \frac {-4( S_3 +(1+a)S_1)^2}{(4S_0^2d_0^3(S_3 +(1+a)S_1)^2)} \left[ \left[ S_0^2 \left\{ \tilde{A} d_0^2 + 2(D\tilde{B})d_0^2 + (D^2\tilde{C})d_0^2 - \tilde{C}d_0(D^2d_0) - 2\tilde{B}d_0(Dd_0) - \right. \right. \right. \nonumber\\
& \left. 2(D\tilde{C})d_0(Dd_0) + 2\tilde{C}(Dd_0)^2 +  \alpha^2\tilde{C}d_0^2\right\}+ 2(1-a^2)(i\alpha)\left\{S_0((D^2U)(S_1+S_3) + DU(DS_1 +  \right. \nonumber\\
& \left. \left. DS_3)) -DU(S_1+S_3)DS_0\right\}\tilde{C}d_0^2 + 2i\alpha(1-a^2)DU(S_1+S_3)S_0\left\{(\tilde{B} + D\tilde{C})d_0^2 - \tilde{C}d_0 (Dd_0)\right\}  \right]\nonumber\\
& + d_0^3 \left[2S_0(S_3 +(1+a)S_1) \left\{ (2(S_1+S_3)(S_2 - (1+a)S_1) + 1 - 2(S_3 + (1+a)S_1)(S_1+ S_3) \right. \right. \nonumber\\
&  2a i\alpha(1-a)(\tau_0^{11}) - (2(S_1+S_3)(S_2 - (1+a)S_1) + 1 + 2(S_1+S_3)(S_3 + (1+a)S_1) )2a(i\alpha) \nonumber\\
& (1+a)\tau_0^{22} \!-\!\left\{(2(S_1 \!\!+\!\! S_3)(S_2\!\! -\!\! (1\!\!+\!\!a)S_1)+1)a + 2(S_1+S_3)(S_3+(1+a)S_1))\right\}\frac{4i\alpha \nu}{ReE} \!- \!4(1\!-\!a^2) \nonumber\\
& \left. (S_1+S_3)(S_3+(1+a)S_1)(D\tau_0^{12})\right\} -8(S_3+(1+a)S_1)^2(1-a^2)\tau_0^{12}(S_0(DS_1+DS_3) - (DS_0)(S_1+\nonumber\\
&\left. \left. S_3))\right] \right],
\label{eqn:c2}
\end{align}
\begin{align}
c_3 &= - \frac{1}{(S_0d_0^2)}\left[ S_0\left\{(\tilde{B})d_0 + 2(D\tilde{C})d_0 - 2(\tilde{C})(Dd_0)\right\} + 2(1-a^2)(i\alpha)(DU)(S_1+S_3)(\tilde{C})d_0 + 2d_0^2 \right. \nonumber\\
& \left. (S_1+S_3)(1-a^2)(\tau_0^{12}) \right], 
\label{eqn:c3} \\
\nonumber \\
c_4 &= - \frac{(\tilde{C})}{(d_0)}.
\label{eqn:c4}
\end{align}
\label{eqn:OSE_ci}
\eseq
where the following coefficients are used in the definition of $c_i$'s above,
\begin{align}
\tilde{A} &= \left( \frac{(1-a)}{2}(DU) + \varepsilon \tau_0^{12} \right) (- \frac{A}{H}) + (\varepsilon \tau_0^{12} - \frac{(1+a)}{2}(DU))(F_1 + \frac{(A)(G)}{H}) - (i\alpha)(D\tau_0^{12}) + \frac{(1-a)}{2}\nonumber\\
& \alpha^2\tau_0^{22} - \frac{(1+a)}{2}\alpha^2\tau_0^{11} - \frac{\alpha^2 \nu}{ReE}, \nonumber \\
\tilde{B} &= \left( \frac{(1-a)}{2}(DU) + \varepsilon \tau_0^{12} \right) (- \frac{B}{H}) + (\varepsilon \tau_0^{12} - \frac{(1+a)}{2}(DU))(F_2 + \frac{(B)(G)}{H}), \nonumber\\
\tilde{C} &= \left( \frac{(1-a)}{2}(DU) + \varepsilon \tau_0^{12} \right) ( \frac{C}{H}) - (\varepsilon \tau_0^{12} - \frac{(1+a)}{2}(DU))(\frac{(C)(G)}{H}) + \frac{(1-a)}{2}\tau_0^{11} - \frac{(1+a)}{2}\tau_0^{22} \nonumber\\
& - \frac{\nu}{ReE}, \nonumber\\
A &= \frac{2(1-a^2)}{(\varepsilon \tau_0^{11}(1-a) + ( i\alpha(U-c) + \frac{\nu}{ReE} + \varepsilon(\tau_0^{11} + 2\tau_0^{22}))(1+a))}\left[(i\alpha(D\tau_0^{22}) + (1+a)\alpha^2\tau_0^{12})\varepsilon \tau_0^{11} - \right. \nonumber\\
& \left. (i\alpha(D\tau_0^{11}) - (1-a)\alpha^2\tau_0^{12})(i\alpha(U - c) + \frac{\nu}{ReE} + \varepsilon(\tau_0^{11} + 2\tau_0^{22})) \right], \nonumber\\
B &= \frac{-4i\alpha(1-a^2)}{(\varepsilon \tau_0^{11}(1\!\!-\!\!a) \!\!+\!\! ( i\alpha(U\!\!-\!\!c)\!\! +\!\! \frac{\nu}{ReE} \!\!+\!\! \varepsilon(\tau_0^{11}\!\! +\!\! 2\tau_0^{22}))(1\!\!+\!\!a))} \big[ (\varepsilon \tau_0^{11})(a\tau_0^{22} + \frac{\nu}{ReE}) + (a\tau_0^{11} + \frac{\nu}{ReE}) \nonumber\\
& \left. (i\alpha(U - c) + \frac{\nu}{ReE} + \varepsilon(\tau_0^{11} + 2\tau_0^{22})) \right], \nonumber \\
C &=  2(1-a^2) \tau_0^{12}, \nonumber \\
H &= \frac{2(1-a^2)}{(\varepsilon \tau_0^{11}(1-a) + ( i\alpha(U-c) + \frac{\nu}{ReE} + \varepsilon(\tau_0^{11} + 2\tau_0^{22}))(1+a))}\!\! \left[ (i\alpha(U\!\!-\!\!c)\!\!+\!\! \frac{\nu}{ReE})^2\!\!+\!\! 6\varepsilon((i\alpha)  \right. \nonumber\\
& \left. (U-c) + \frac{\nu}{ReE})(\tau_0^{11} + 2\tau_0^{22}) + 2(\varepsilon)^2 (\tau_0^{11} + \tau_0^{22})^2\right], \nonumber \\
F_1 &= \frac{(i\alpha((1+a)(D\tau_0^{22}) + (1-a)(D\tau_0^{11})) + 4a\alpha^2\tau_0^{12})}{(\varepsilon \tau_0^{11}(1-a) + ( i\alpha(U-c) + \frac{\nu}{ReE} + \varepsilon(\tau_0^{11} + 2\tau_0^{22}))(1+a))}, \nonumber \\
F_2 &= \frac{(2ai\alpha((1-a)\tau_0^{11} - (1+a)\tau_0^{22} - \frac{2\nu}{ReE}))}{(\varepsilon \tau_0^{11}(1-a) + ( i\alpha(U-c) + \frac{\nu}{ReE} + \varepsilon(\tau_0^{11} + 2\tau_0^{22}))(1+a))}, \nonumber \\
G &= \frac{ ((i\alpha(U-c) +2\varepsilon\tau_0^{11}  + \frac{\nu}{ReE})(1-a) + 2\varepsilon\tau_0^{22})}{(\varepsilon \tau_0^{11}(1-a) + ( i\alpha(U-c) + \frac{\nu}{ReE} + \varepsilon(\tau_0^{11} + 2\tau_0^{22}))(1+a))}, \nonumber \\
d_0 &= \frac{2(1-a^2)(DU)}{H} \left[(\frac{1-a}{2})(DU) + \varepsilon \tau_0^{12}\right]\!\! +\!\! S_1\!\! +\!\! \frac{2(1-a^2)(DU)(G)}{H}\!\! \left[(\frac{1+a}{2})(DU)\!\!-\!\!\varepsilon \tau_0^{12}\right], \nonumber \\
S_1 &= i\alpha(U-c) + \frac{\nu}{ReE} + \varepsilon(\tau_0^{11}+\tau_0^{22}), \nonumber \\
S_2 &= \varepsilon(1-a)\tau_0^{11} - \varepsilon(1+a)\tau_0^{22}, \nonumber \\
S_3 &= \varepsilon(1-a)\tau_0^{11} + \varepsilon(1+a)\tau_0^{22}, \nonumber \\
S_0 &= aS_1S_2 + S_1S_3 + (1-a^2)S_1^2.
\label{eqn:C_coeff}
\end{align}
%
%\begin{align}
%E &= \frac{2(1-a^2)}{(\varepsilon \tau_0^{11}(1-a) + ( i\alpha(U-c) + \frac{1}{We} + \varepsilon(\tau_0^{11} + 2\tau_0^{22}))(1+a))} \left[ (i\alpha(U-c)+ \frac{1}{We})^2+ 6\varepsilon((i\alpha)  \right. \nonumber\\
%& \left. (U-c) + \frac{1}{We})(\tau_0^{11} + 2\tau_0^{22}) + 2(\varepsilon)^2 (\tau_0^{11} + \tau_0^{22})^2\right], \nonumber \\
%
%
%F_1 &= \frac{(i\alpha((1+a)(D\tau_0^{22}) + (1-a)(D\tau_0^{11})) + 4a\alpha^2\tau_0^{12})}{(\varepsilon \tau_0^{11}(1-a) + ( i\alpha(U-c) + \frac{1}{We} + \varepsilon(\tau_0^{11} + 2\tau_0^{22}))(1+a))}, \nonumber \\
%
%
%F_2 &= \frac{(2ai\alpha((1-a)\tau_0^{11} - (1+a)\tau_0^{22} - \frac{2}{We}))}{(\varepsilon \tau_0^{11}(1-a) + ( i\alpha(U-c) + \frac{1}{We} + \varepsilon(\tau_0^{11} + 2\tau_0^{22}))(1+a))}, \nonumber \\
%
%
%G &= \frac{ ((i\alpha(U-c) +2\varepsilon\tau_0^{11}  + \frac{1}{We})(1-a) + 2\varepsilon\tau_0^{22})}{(\varepsilon \tau_0^{11}(1-a) + ( i\alpha(U-c) + \frac{1}{We} + \varepsilon(\tau_0^{11} + 2\tau_0^{22}))(1+a))}, \nonumber \\
%
%
%d_0 &= \frac{2(1-a^2)(DU)}{E} \left[(\frac{1-a}{2})(DU) + \varepsilon \tau_0^{12}\right] + S_1 + \frac{2(1-a^2)(DU)(G)}{E}\left[(\frac{1+a}{2})(DU)-\varepsilon \tau_0^{12}\right], \nonumber \\
%
%
%S_1 &= i\alpha(U-c) + \frac{1}{We} + \varepsilon(\tau_0^{11}+\tau_0^{22}), \nonumber \\
%
%
%S_2 &= \varepsilon(1-a)\tau_0^{11} - \varepsilon(1+a)\tau_0^{22}, \nonumber \\
%
%
%S_3 &= \varepsilon(1-a)\tau_0^{11} + \varepsilon(1+a)\tau_0^{22}, \nonumber \\
%
%
%S_0 &= aS_1S_2 + S_1S_3 + (1-a^2)S_1^2.
%\label{eqn:C_coeff}
%\end{align}
%
%
%
along with the first derivatives,
\begin{align*}
D\tilde{A} &= ((\frac{1-a}{2})(D^2U) + \varepsilon(D\tau_0^{12}))(-\frac{A}{H}) + ((\frac{1-a}{2})(DU) + \varepsilon\tau_0^{12})(\frac{(DH)A - H(DA)}{H^2})+ (\varepsilon (D\tau_0^{12})\nonumber\\
& -(\frac{1+a}{2})(D^2U))(F_1 + \frac{AG}{H})+ (\varepsilon \tau_0^{12}-(\frac{1+a}{2})(DU))(DF_1 + \frac{HA(DG)+HG(DA)-AG(DH)}{H^2}) \nonumber\\
& -i\alpha(D^2\tau_0^{12}) + (\frac{1-a}{2})\alpha^2(D\tau_0^{22}) - (\frac{1+a}{2})\alpha^2(D\tau_0^{11}), \nonumber \\
D\tilde{B} &= ((\frac{1-a}{2})(D^2U) + \varepsilon(D\tau_0^{12}))(-\frac{B}{H}) + ((\frac{1-a}{2})(DU) + \varepsilon\tau_0^{12})(\frac{(DH)B - H(DB)}{H^2})+ (\varepsilon (D\tau_0^{12})\nonumber\\
& -(\frac{1+a}{2})(D^2U))(F_2 + \frac{BG}{H})+ (\varepsilon \tau_0^{12}-(\frac{1+a}{2})(DU))(DF_2 + \frac{HB(DG)+HG(DB)-BG(DH)}{H^2}), \nonumber \\
D\tilde{C} &= ((\frac{1-a}{2})(D^2U) + \varepsilon(D\tau_0^{12}))(\frac{C}{H}) + ((\frac{1-a}{2})(DU) + \varepsilon\tau_0^{12})(\frac{(DC)H - C(DH)}{H^2})- (\varepsilon (D\tau_0^{12})\nonumber\\
& -(\frac{1+a}{2})(D^2U))(\frac{CG}{H})- (\varepsilon \tau_0^{12} - (\frac{1+a}{2})(DU))(\frac{HC(DG)+HG(DC)-CG(DH)}{H^2})+ (\frac{1-a}{2}) \nonumber\\ 
& (D\tau_0^{11}) -(\frac{1+a}{2})(D\tau_0^{22}), \nonumber\\
DA &= \frac{2(1-a^2)}{(\varepsilon \tau_0^{11}(1-a) + ( i\alpha(U-c) + \frac{\nu}{ReE} + \varepsilon(\tau_0^{11} + 2\tau_0^{22}))(1+a))^2}\left[(\varepsilon \tau_0^{11}(1-a) + ( i\alpha(U-c) + \right. \nonumber\\
& \frac{\nu}{ReE} + \varepsilon(\tau_0^{11} + 2\tau_0^{22}))(1+a))\left\{(i\alpha D^2\tau_0^{22}+(1+a)\alpha^2D\tau_0^{12})(\varepsilon \tau_0^{11})+(i\alpha D\tau_0^{22}+(1+a)\alpha^2\tau_0^{12}) \right.\nonumber\\
& (\varepsilon D\tau_0^{11})-(i\alpha D^2\tau_0^{11}\!\!-\!\!(1-a)\alpha^2D\tau_0^{12})(i\alpha(U\!\!-\!\!c) \!\!+\!\! \frac{\nu}{ReE} + \varepsilon(\tau_0^{11} + 2\tau_0^{22})) - (i\alpha D\tau_0^{11}-(1-a)\nonumber\\
& \left. \alpha^2\tau_0^{12})(i\alpha(DU) + \varepsilon(D\tau_0^{11} + 2D\tau_0^{22})\right\} - \left\{(i\alpha D\tau_0^{22}+(1+a)\alpha^2\tau_0^{12})(\varepsilon\tau_0^{11}) - (i\alpha D\tau_0^{11}-(1-a) \right.\nonumber\\
& \left.\left. \alpha^2\tau_0^{12})(i\alpha(U-c) + \frac{\nu}{ReE} + \varepsilon(\tau_0^{11} + 2\tau_0^{22}))\right\}(2\varepsilon D\tau_0^{11}+(1+a)(i\alpha(DU)+2\varepsilon D\tau_0^{22}))\right], \nonumber \\
DB &= \frac{-4i\alpha(1-a^2)}{(\varepsilon \tau_0^{11}(1-a) + ( i\alpha(U-c) + \frac{\nu}{ReE} + \varepsilon(\tau_0^{11} + 2\tau_0^{22}))(1+a))^2}\left[(\varepsilon \tau_0^{11}(1-a) + ( i\alpha(U-c) + \right.\nonumber\\
& \frac{\nu}{ReE} + \varepsilon(\tau_0^{11} + 2\tau_0^{22}))(1+a))\left\{\varepsilon D\tau_0^{11}(a\tau_0^{22}+\frac{\nu}{ReE})+a\varepsilon\tau_0^{11} D\tau_0^{22} + (aD\tau_0^{11})(i\alpha(U-c) + \frac{\nu}{ReE} \right.\nonumber\\
& \left. + \varepsilon(\tau_0^{11}+2\tau_0^{22}))+(a\tau_0^{11}+\frac{\nu}{ReE})(i\alpha(DU)+\varepsilon(D\tau_0^{11} + 2D\tau_0^{22}))\right\}-(2\varepsilon D\tau_0^{11} + ( i\alpha (DU) + 2\varepsilon \nonumber\\
& \left. D\tau_0^{22})(1+a))\left\{\varepsilon \tau_0^{11}(a\tau_0^{22}+\frac{\nu}{ReE}) + (a\tau_0^{11} + \frac{\nu}{ReE})(i\alpha(U-c) + \frac{\nu}{ReE}+ \varepsilon(\tau_0^{11}+2\tau_0^{22}))\right\} \right], \nonumber \\
DC &= 2(1-a^2)(D\tau_0^{12}), \nonumber \\
DH &= \frac{2(1-a^2)}{(\varepsilon \tau_0^{11}(1-a) + ( i\alpha(U-c) + \frac{\nu}{ReE} + \varepsilon(\tau_0^{11} + 2\tau_0^{22}))(1+a))^2}\left[(\varepsilon \tau_0^{11}(1-a) + ( i\alpha(U-c) + \right. \nonumber\\
& \frac{\nu}{ReE} + \varepsilon(\tau_0^{11} + 2\tau_0^{22}))(1+a))\left\{2i\alpha(DU)(i\alpha(U-c) + \frac{\nu}{ReE})+6\varepsilon i\alpha(DU)(\tau_0^{11}+\tau_0^{22})+6\varepsilon (i\alpha \right. \nonumber\\
& \left. (U-c)+\frac{\nu}{ReE})(D\tau_0^{11}+D\tau_0^{22}) + (4\varepsilon^2)(\tau_0^{11}+\tau_0^{22})(D\tau_0^{11}+D\tau_0^{22})\right\}-(2\varepsilon (D\tau_0^{11})+(i\alpha(DU)+ \nonumber\\
& \left. 2\varepsilon(D\tau_0^{22})(1\!+\!a))\!\!\left\{(i\alpha(U\!\!-\!\!c)+\frac{\nu}{ReE})^2\!\!+\!\!6\varepsilon(i\alpha(U\!\!-\!\!c)+\frac{\nu}{ReE})(\tau_0^{11}\!+\!\tau_0^{22})+2\varepsilon^2(\tau_0^{11}+\tau_0^{22})^2\right\}  \right],\nonumber
\end{align*}
\begin{align}
DF_1 &= \frac{1}{(\varepsilon \tau_0^{11}(1-a) + ( i\alpha(U-c) + \frac{\nu}{ReE} + \varepsilon(\tau_0^{11} + 2\tau_0^{22}))(1+a))^2}\left[(\varepsilon \tau_0^{11}(1-a) +( i\alpha(U-c) + \right. \nonumber\\
&  \frac{\nu}{ReE} + \varepsilon(\tau_0^{11} + 2\tau_0^{22}))(1+a))(i\alpha((1+a)D^2\tau_0^{22}+(1-a)D^2\tau_0^{11})+4a\alpha^2(D\tau_0^{12}))-(2\varepsilon(D\tau_0^{11}) \nonumber\\
& \left. +(i\alpha(DU)+2\varepsilon(D\tau_0^{22})(1+a))(i\alpha((1+a)D\tau_0^{22}+(1-a)D\tau_0^{11})+4a\alpha^2\tau_0^{12})\right], \nonumber \\
DF_2 &= \frac{2ai\alpha}{(\varepsilon \tau_0^{11}(1-a) + ( i\alpha(U-c) + \frac{\nu}{ReE} + \varepsilon(\tau_0^{11} + 2\tau_0^{22}))(1+a))^2}\left[(\varepsilon \tau_0^{11}(1-a) + ( i\alpha(U-c) + \right. \nonumber\\
& \frac{\nu}{ReE} + \varepsilon(\tau_0^{11} + 2\tau_0^{22}))(1+a))((1-a)(D\tau_0^{11})-(1+a)(D\tau_0^{22})) - (2\varepsilon(D\tau_0^{11})+(i\alpha(DU)+2\varepsilon \nonumber\\
& \left. (D\tau_0^{22}))(1+a))((1-a)\tau_0^{11}-(1+a)\tau_0^{22}- \frac{2\nu}{ReE})\right], \nonumber \\%
DG &= \frac{1}{(\varepsilon \tau_0^{11}(1-a) + ( i\alpha(U-c) + \frac{\nu}{ReE} + \varepsilon(\tau_0^{11} + 2\tau_0^{22}))(1+a))^2}\left[(\varepsilon \tau_0^{11}(1-a) + ( i\alpha(U-c) + \right.\nonumber\\
& \frac{\nu}{ReE} + \varepsilon(\tau_0^{11} + 2\tau_0^{22}))(1+a))((i\alpha(DU)+2\varepsilon(D\tau_0^{11}))(1-a)+2\varepsilon(D\tau_0^{22})) - ((i\alpha(DU)+2\varepsilon \nonumber\\
& \left. (D\tau_0^{22}))(1+a)+2\varepsilon(D\tau_0^{11}))((i\alpha(U-c)+\frac{\nu}{ReE}+2\varepsilon\tau_0^{11})(1-a)+2\varepsilon\tau_0^{22})\right], \nonumber \\
Dd_0 &= DS_1 + 2(1-a^2)\left[((\frac{1-a}{2})(D^2U)+\varepsilon(D\tau_0^{12}))(\frac{(DU)}{H})+((\frac{1-a}{2})(DU)+\varepsilon(\tau_0^{12})) \right. \nonumber\\
& (\frac{(H(D^2U)-(DU)(DH))}{H^2})-((\varepsilon D\tau_0^{12}-(\frac{1+a}{2})(D^2U))(\frac{(DU)G}{H})+(\varepsilon \tau_0^{12}-(\frac{1+a}{2})(DU))\nonumber\\
& \left. (\frac{H(D^2U)G + H(DU)(DG)-(DH)(DU)G}{H^2})\right], \nonumber \\
DS_1 &= i\alpha(DU) + \varepsilon(D\tau_0^{11}+D\tau_0^{22}), \nonumber \\
DS_2 &= \varepsilon(1-a)D\tau_0^{11} - \varepsilon(1+a)D\tau_0^{22}, \nonumber \\
DS_3 &= \varepsilon(1-a)D\tau_0^{11} + \varepsilon(1+a)D\tau_0^{22}, \nonumber \\
DS_0 &= a(DS_1)S_2 + aS_1(DS_2) + (DS_1)S_3 + S_1(DS_3) + 2(1-a^2)S_1(DS_1).
\end{align}

\noindent and the second derivatives,

\begin{align*}
D^2\tilde{A} &= ((\frac{1-a}{2})(D^3U) + \varepsilon(D^2\tau_0^{12}))(-\frac{A}{H}) + 2((\frac{1-a}{2})(D^2U)+\varepsilon (D\tau_0^{12}))(\frac{(DH)A-H(DA)}{H^2})+((\frac{1-a}{2}) \nonumber\\
& (DU)+\varepsilon \tau_0^{12})\frac{H((D^2H)A-H(D^2A))-2(DH)((DH)A-H(DA))}{H^3} + (\varepsilon D^2\tau_0^{12} - (\frac{1+a}{2})(D^3U))(F_1+\nonumber\\
& \frac{AG}{H}) + 2(\varepsilon D\tau_0^{12} - (\frac{1+a}{2})(D^2U))(DF_1 + \frac{HA(DG)+HG(DA)-AG(DH)}{H^2})+ (\varepsilon \tau_0^{12}-(\frac{1+a}{2})(DU)) \nonumber\\
& (D^2F_1 + \frac{1}{H^3}(H^2(A(D^2G)+2(DG)(DH)+G(D^2A))-AGH(D^2H))-2(DH)(H(A(DG)+G(DA))-\nonumber\\
& AG(DH)))- i\alpha D^3\tau_0^{12}+(\frac{1-a}{2})\alpha^2D^2\tau_0^{22}-(\frac{1+a}{2})\alpha^2D^2\tau_0^{11}, 
\end{align*}

\begin{align*}
D^2\tilde{B} &= ((\frac{1-a}{2})(D^3U) + \varepsilon(D^2\tau_0^{12}))(-\frac{B}{H}) + 2((\frac{1-a}{2})(D^2U) + \varepsilon D\tau_0^{12})(\frac{(DH)B - H(DB)}{H^2}) \nonumber\\
& + ((\frac{1-a}{2})(DU) + \varepsilon \tau_0^{12})(\frac{H(D^2H)B - H^2(D^2B) - 2(DH)(B(DH)-H(DB))}{H^3}) + (\varepsilon (D^2\tau_0^{12})\nonumber\\
& - (\frac{1+a}{2})(D^3U))(F_2 + \frac{BG}{H})+ 2(\varepsilon D\tau_0^{12}-(\frac{1+a}{2})(D^2U))(DF_2 + \frac{HB(DG)+HG(DB)-BG(DH)}{H^2}) \nonumber\\
& + (\varepsilon \tau_0^{12}-(\frac{1+a}{2})(DU))(D^2F_2 + \frac{1}{H^3}(H^2B(D^2G)+2H^2(DG)(DB)+H^2G(D^2B)-BGH(D^2H)\nonumber\\
& -2(DH)(HB(DG)+HG(DB)-BG(DH)))), \nonumber \\
D^2\tilde{C} &= ((\frac{1-a}{2})(D^3U) + \varepsilon(D^2\tau_0^{12}))(\frac{C}{H}) + 2((\frac{1-a}{2})(D^2U)+\varepsilon (D\tau_0^{12}))(\frac{(DC)H-C(DH)}{H^2})+((\frac{1-a}{2}) \nonumber\\
& (DU)+\varepsilon \tau_0^{12})\frac{H^2(D^2C)-CH(D^2H)-2(DH)((DC)H-C(DH))}{H^3} - (\varepsilon D^2\tau_0^{12} - (\frac{1+a}{2})(D^3U))(\frac{CG}{H})\nonumber\\
& - 2(\varepsilon D\tau_0^{12} - (\frac{1+a}{2})D^2U)(\frac{HC(DG)+HG(DC)-CG(DH)}{H^2}) - (\varepsilon \tau_0^{12}-DU(\frac{1+a}{2}))  \nonumber\\
& (\frac{H^2(C(D^2G)+2(DG)(DC)+G(D^2C))-CGH(D^2H)-2(DH)(HC(DG)+HG(DC)-CG(DH))}{H^3})\nonumber\\
& +(\frac{1-a}{2})D^2\tau_0^{11}-(\frac{1+a}{2})D^2\tau_0^{22}, \nonumber \\
D^2G &= \frac{1}{(\varepsilon \tau_0^{11}(1-a)+(i\alpha(U-c)+\frac{\nu}{ReE}+\varepsilon(\tau_0^{11}+2\tau_0^{22}))(1+a))^3}\left[(\varepsilon \tau_0^{11}(1-a)+(i\alpha(U-c)+\right. \nonumber\\
& \frac{\nu}{ReE}+\varepsilon(\tau_0^{11}+2\tau_0^{22}))(1+a))\left\{(\varepsilon \tau_0^{11}(1-a)+(i\alpha(U-c)+\frac{\nu}{ReE}+\varepsilon(\tau_0^{11}+2\tau_0^{22}))(1+a))\right. \nonumber\\
& ((i\alpha(D^2U)+2\varepsilon D^2\tau_0^{11})(1-a)+2\varepsilon D^2\tau_0^{22})-(2\varepsilon D^2\tau_0^{11}+(i\alpha(D^2U)+2\varepsilon D^2\tau_0^{22})(1+a))\nonumber\\
& \left. ((i\alpha(U-c)+\frac{\nu}{ReE}+2\varepsilon \tau_0^{11})(1-a)+2\varepsilon \tau_0^{22})\right\}-2(2\varepsilon D\tau_0^{11}+(i\alpha(DU)+2\varepsilon D\tau_0^{22})(1+a))\nonumber\\
& \left\{(\varepsilon \tau_0^{11}(1-a)+(i\alpha(U-c)+\frac{\nu}{ReE}+\varepsilon(\tau_0^{11}+2\tau_0^{22}))(1+a))(2\varepsilon D\tau_0^{22}+(i\alpha(DU)+2\varepsilon D\tau_0^{11})\right.\nonumber\\
& \left. \left. (1-a))-(2\varepsilon D\tau_0^{11}+(i\alpha(DU)+2\varepsilon D\tau_0^{22})(1+a))((i\alpha(U-c)+\frac{\nu}{ReE}+2\varepsilon \tau_0^{11})(1-a)+2\varepsilon \tau_0^{22})\right\}\right], \nonumber \\
D^2F_1 &= \frac{1}{(\varepsilon \tau_0^{11}(1-a)+(i\alpha(U-c)+\frac{\nu}{ReE}+\varepsilon(\tau_0^{11}+2\tau_0^{22}))(1+a))^3}\left[(\varepsilon \tau_0^{11}(1-a)+(i\alpha(U-c)+\right. \nonumber\\
& \frac{\nu}{ReE}+\varepsilon(\tau_0^{11}+2\tau_0^{22}))(1+a))\left\{(\varepsilon \tau_0^{11}(1-a)+(i\alpha(U-c)+\frac{\nu}{ReE}+\varepsilon(\tau_0^{11}+2\tau_0^{22}))(1+a))\right. \nonumber\\
& (i\alpha((1+a)D^3\tau_0^{22}+(1-a)D^3\tau_0^{11})+4a\alpha^2D^2\tau_0^{12}) \!-\!(2\varepsilon D^2\tau_0^{11}\!+\!(i\alpha(D^2U)+2\varepsilon D^2\tau_0^{22})(1+a))\nonumber\\
& \left. (i\alpha((1+a)D\tau_0^{22}+(1-a)D\tau_0^{11})+4a\alpha^2D\tau_0^{12})\right\}-2(2\varepsilon D\tau_0^{11}+(i\alpha(DU)+2\varepsilon D\tau_0^{22})(1+a))\left\{(\varepsilon \right.\nonumber\\
& \tau_0^{11}(1-a)+(i\alpha(U-c)+\frac{\nu}{ReE}+\varepsilon(\tau_0^{11}+2\tau_0^{22}))(1+a))(i\alpha((1+a)D^2\tau_0^{22}+(1-a)D^2\tau_0^{11})+4a \nonumber\\
& \left.\left. \alpha^2D\tau_0^{12})-(2\varepsilon D\tau_0^{11}+(i\alpha(DU)+2\varepsilon D\tau_0^{22})(1+a))(i\alpha((1+a)D\tau_0^{22}+(1-a)D\tau_0^{11})+4a\alpha^2\tau_0^{12})\right\}\right], \nonumber\\
D^2C &= 2(1-a^2)D^2\tau_0^{12}, \nonumber %\\
\end{align*}
\begin{align*}
D^2A &=\frac{2(1-a^2)}{(\varepsilon \tau_0^{11}(1-a) + ( i\alpha(U-c) + \frac{\nu}{ReE} + \varepsilon(\tau_0^{11} + 2\tau_0^{22}))(1+a))^3}\left[(\varepsilon \tau_0^{11}(1-a) + ( i\alpha(U-c) + \frac{\nu}{ReE} \right.\nonumber\\
& + \varepsilon(\tau_0^{11} + 2\tau_0^{22}))(1+a))\left\{(2\varepsilon D\tau_0^{11}+(i\alpha(DU)+2\varepsilon(D\tau_0^{22}))(1+a))((i\alpha D^2\tau_0^{22}+(1+a)\alpha^2D\tau_0^{12}) \right. \nonumber\\
&  (\varepsilon \tau_0^{11})+(i\alpha D\tau_0^{22}+(1+a)\alpha^2\tau_0^{12})(\varepsilon D\tau_0^{11}) - (i\alpha D^2\tau_0^{11} - (1-a)\alpha^2D\tau_0^{12})(i\alpha(U-c) + \frac{\nu}{ReE}+\varepsilon \nonumber\\
& (\tau_0^{11}+2\tau_0^{22}))-(i\alpha D\tau_0^{11}-(1-a)\alpha^2\tau_0^{12})(i\alpha(DU)+\varepsilon(D\tau_0^{11}+2D\tau_0^{22}))) +(\varepsilon \tau_0^{11}(1-a)+(i\alpha(U\nonumber\\
& -c)\!+\! \frac{\nu}{ReE}+\varepsilon(\tau_0^{11}\!+\!2\tau_0^{22}))(1+a))((i\alpha D^3\tau_0^{22}+(1+a)\alpha^2D^2\tau_0^{12})(\varepsilon \tau_0^{11})+2(i\alpha D^2\tau_0^{22}\!+\!(1\!+\!a)\alpha^2\nonumber\\
& D\tau_0^{12})(\varepsilon D\tau_0^{11})+(i\alpha D\tau_0^{22}+(1+a)\alpha^2\tau_0^{12})(\varepsilon D^2\tau_0^{11})-(i\alpha D^3\tau_0^{11}-(1-a)\alpha^2D^2\tau_0^{12})(i\alpha(U-c)+\nonumber\\
& \frac{\nu}{ReE}+\varepsilon(\tau_0^{11}+2\tau_0^{22}))-2(i\alpha D^2\tau_0^{11}-(1-a)\alpha^2D\tau_0^{12})(i\alpha(DU)+\varepsilon(D\tau_0^{11}+2D\tau_0^{22}))-(i\alpha D\tau_0^{11}- \nonumber\\
& (1-a)\alpha^2\tau_0^{12})(i\alpha(D^2U)+\varepsilon(D^2\tau_0^{11}+2D^2\tau_0^{22}))- ((i\alpha D^2\tau_0^{22}+(1+a)\alpha^2D\tau_0^{12})(\varepsilon \tau_0^{11})+(i\alpha D\tau_0^{22}\nonumber\\
& +(1+a)\alpha^2\tau_0^{12})(\varepsilon D\tau_0^{11})-(i\alpha D^2\tau_0^{11} - (1-a)\alpha^2D\tau_0^{12})(i\alpha(U-c)+ \frac{\nu}{ReE}+\varepsilon(\tau_0^{11}+2\tau_0^{22}))-(i\alpha \nonumber\\
& D\tau_0^{11}-(1-a)\alpha^2\tau_0^{12})(i\alpha(DU)+\varepsilon(D\tau_0^{11}+2D\tau_0^{22})))(2\varepsilon D\tau_0^{11}+(1+a)(i\alpha(DU)+2\varepsilon D\tau_0^{22}))-\nonumber\\
& ((i\alpha D\tau_0^{22}+(1+a)\alpha^2\tau_0^{12})(\varepsilon \tau_0^{11})-(i\alpha D\tau_0^{11}-(1-a)\alpha^2\tau_0^{12})(i\alpha(U-c)+\frac{\nu}{ReE}+\varepsilon(\tau_0^{11}+2\tau_0^{22})))\nonumber\\
& \left. (2\varepsilon D^2\tau_0^{11}+(1+a)(i\alpha(D^2U)+2\varepsilon D^2\tau_0^{22}))\right\} - 2(2\varepsilon D\tau_0^{11}+(i\alpha(DU)+2\varepsilon D\tau_0^{22})(1+a))\left\{((i\alpha \right.\nonumber\\
& D^2\tau_0^{22}+(1+a)\alpha^2D\tau_0^{12})(\varepsilon \tau_0^{11})+(i\alpha D\tau_0^{22} +(1+a)\alpha^2\tau_0^{12})(\varepsilon D\tau_0^{11})-(i\alpha D^2\tau_0^{11}-(1-a)\alpha^2 D\tau_0^{12})\nonumber\\
& (i\alpha(U\!\!-\!\!c)+\frac{\nu}{ReE}+\varepsilon(\tau_0^{11}+2\tau_0^{22}))\!-\!(i\alpha D\tau_0^{11}-(1-a)\alpha^2\tau_0^{12})(i\alpha(DU)+\varepsilon(D\tau_0^{11}+2D\tau_0^{22})))(\varepsilon \tau_0^{11} \nonumber\\
&(1-a)+(i\alpha(U\!\!-\!\!c)+\frac{\nu}{ReE}+\varepsilon(\tau_0^{11}+2\tau_0^{22}))(1+a)) \!\!-\!\! ((i\alpha D\tau_0^{22}+(1+a)\alpha^2\tau_0^{12})(\varepsilon \tau_0^{11})-(i\alpha D\tau_0^{11}\nonumber\\
& \left. \left. -(1-a)\alpha^2\tau_0^{12})(i\alpha(U-c)+\frac{\nu}{ReE}+\varepsilon(\tau_0^{11}+2\tau_0^{22})))(2\varepsilon D\tau_0^{11} + (1+a)(i\alpha(DU)+2\varepsilon D\tau_0^{22}))\right\}\right],\nonumber\\
D^2H &= \frac{2(1-a^2)}{(\varepsilon \tau_0^{11}(1-a) + ( i\alpha(U-c) + \frac{\nu}{ReE} + \varepsilon(\tau_0^{11} + 2\tau_0^{22}))(1+a))^3}\left[(\varepsilon \tau_0^{11}(1-a) + ( i\alpha(U-c) + \frac{\nu}{ReE} \right. \nonumber\\
& + \varepsilon(\tau_0^{11} \!+\! 2\tau_0^{22}))(1\!+\!a))\left\{(\varepsilon \tau_0^{11}(1\!-\!a) + ( i\alpha(U\!-\!c) +\frac{\nu}{ReE} + \varepsilon(\tau_0^{11} + 2\tau_0^{22}))(1+a))(2(i\alpha(U-c) \right. \nonumber\\
& +\frac{\nu}{ReE})i\alpha(D^2U)\!-\!2\alpha^2(DU)^2\!+\!6\varepsilon i\alpha(D^2U)(\tau_0^{11}\!+\!\tau_0^{22})\!+\!12\varepsilon i\alpha(DU)(D\tau_0^{11}\!+\!D\tau_0^{22})+\!6\!\varepsilon(i\alpha(U\!-\!c)\nonumber\\
& +\frac{\nu}{ReE})(D^2\tau_0^{11}\!\!+\!\!D^2\tau_0^{22})\!+\!4\varepsilon^2(D\tau_0^{11}\!\!+\!\!D\tau_0^{22})^2\!+\!4\varepsilon^2(\tau_0^{11}\!\!+\!\!\tau_0^{22})\!\!\! \left. (D^2\tau_0^{11}\!\!+\!\!D^2\tau_0^{22})\right\}\!-\!(2\varepsilon (D^2\tau_0^{11}\!+\!D^2\tau_0^{22})\nonumber\\
&+i\alpha(D^2U)(1\!\!+\!\!a))((i\alpha(U\!\!-\!\!c)\!\!+\!\!\frac{\nu}{ReE})^2 \!\!+\!\!6 \varepsilon (i\alpha(U-c)+\frac{\nu}{ReE})(\tau_0^{11} + \tau_0^{22})+2\varepsilon^2(\tau_0^{11}+\tau_0^{22})^2) -2(2\varepsilon\nonumber\\
& (D\tau_0^{11}\!\!+\!\!D\tau_0^{22})+i\alpha DU )(1\!\!+\!\!a)) \left\{(\varepsilon \tau_0^{11}(1\!\!-\!\!a)\!\! +\!\! ( i\alpha(U\!\!-\!\!c) + \frac{\nu}{ReE} + \varepsilon(\tau_0^{11} + 2\tau_0^{22})) (1+a))(2(i\alpha DU \right.\nonumber\\
& (i\alpha(U-c)+\frac{\nu}{ReE})+ 6\varepsilon i \alpha DU (\tau_0^{11}+\tau_0^{22}) + 6\varepsilon (i\alpha (U-c)+\frac{\nu}{ReE})(D\tau_0^{11}+D\tau_0^{22}) +4\varepsilon^2(\tau_0^{11}+\tau_0^{22}) \nonumber\\
& \left. (D\tau_0^{11}\!\!+\!\!D\tau_0^{22})\right\} \!\!-\!\!(2\varepsilon D\tau_0^{11}\!\!+\!\! (i\alpha(DU)+2\varepsilon D\tau_0^{22}) (1+a))((i\alpha(U\!\!-\!\!c)+ \frac{\nu}{ReE})^2 + 6\varepsilon(i\alpha(U\!\!-\!\!c) +\frac{\nu}{ReE}) \nonumber\\
& \left. \left. (\tau_0^{11}+ \tau_0^{22}) + 2\varepsilon^2 (\tau_0^{11} + \tau_0^{22})^2)\right\}\right], \nonumber
\end{align*}

\begin{align}
D^2F_2 &=  \frac{2ai\alpha}{(\varepsilon \tau_0^{11}(1-a) + ( i\alpha(U-c) + \frac{\nu}{ReE} + \varepsilon(\tau_0^{11} + 2\tau_0^{22}))(1+a))^3}\left[(\varepsilon \tau_0^{11}(1-a) + ( i\alpha(U-c) +\frac{\nu}{ReE} \right. \nonumber\\
& + \varepsilon(\tau_0^{11}\!\! +\!\! 2\tau_0^{22}))(1+a))((\varepsilon \tau_0^{11}(1-a) \!\!+\!\! ( i\alpha(U-c) +\frac{\nu}{ReE} + \varepsilon(\tau_0^{11}\!\!+\!\! 2\tau_0^{22}))(1+a))((1-a)(D^2\tau_0^{11})\nonumber\\
& -(1+a)(D^2\tau_0^{22}))\!\! -\!\! (2\varepsilon(D^2\tau_0^{11})+(i\alpha(D^2U)+2\varepsilon (D^2\tau_0^{22}))(1+a))((1-a)\tau_0^{11}\!\!-\!\!(1+a)\tau_0^{22}\!-\! \frac{2\nu}{ReE})) \nonumber\\
& - 2(2\varepsilon D\tau_0^{11} \!\!+\!\! ( i\alpha DU + 2\varepsilon D\tau_0^{22})(1\!\!+\!\!a))((\varepsilon \tau_0^{11}(1-a) + ( i\alpha(U-c) + \frac{\nu}{ReE} + \varepsilon(\tau_0^{11} + 2\tau_0^{22}))(1+a)) \nonumber\\
& \left. ((1-a)D\tau_0^{11}\!\!-\!\!(1\!\!+\!\!a)D\tau_0^{22})\!\! -\!\! (2\varepsilon D\tau_0^{11}+(i\alpha DU\!\!+\!\!2\varepsilon D\tau_0^{22})(1+a))((1-a)\tau_0^{11}\!\! -\!\!(1+a)\tau_0^{22} - \frac{2\nu}{ReE})) \right],\nonumber\\
%& \left. \nonumber \\
%
D^2d_0 &= D^2S_1 + 2(1-a^2)\left[((\frac{1-a}{2})(D^3U)+\varepsilon(D^2\tau_0^{12}))(\frac{(DU)}{H})+2((\frac{1-a}{2})(D^2U)+\varepsilon(D\tau_0^{12})) \right.\nonumber\\
& (\frac{(H(D^2U)-(DU)(DH))}{H^2})+(\frac{H(H(D^3U)-(DU)(D^2H))-2(DH)(H(D^2U)-(DH)(DU))}{H^3})\nonumber\\
& ((\frac{1-a}{2})(DU)+\varepsilon \tau_0^{12})-\left\{((\varepsilon D^2\tau_0^{12}-(\frac{1+a}{2})(D^3U))(\frac{(DU)G}{H})+2(\varepsilon D\tau_0^{12}-(\frac{1+a}{2})(D^2U))\right. \nonumber\\
& (\frac{H(D^2U)G + H(DU)(DG)-(DH)(DU)G}{H^2})+(\varepsilon \tau_0^{12}-(\frac{1+a}{2})(DU))\frac{1}{H^3}(H(H(D^3U)G + 2H(D^2U) \nonumber\\
& \left. \left. (DG)+H(DU(D^2G) -G(DU)(D^2H)-2(DH)(H(D^2U)G+H(DU)(DG)-(DU)(DH)G)))\right\}\right], \nonumber \\
D^2B &= \frac{-4i\alpha(1-a^2)}{(\varepsilon\tau_0^{11}(1-a)+(i\alpha(U-c)+\frac{\nu}{ReE}+\varepsilon(\tau_0^{11}+2\tau_0^{22}))(1+a))^3}\left[(\varepsilon\tau_0^{11}(1-a)+(i\alpha(U-c)+ \right. \nonumber\\
& \frac{\nu}{ReE}+\varepsilon(\tau_0^{11}+2\tau_0^{22}))(1+a))((2\varepsilon D\tau_0^{11}+(i\alpha(DU)+2D\tau_0^{22})(1+a))(\varepsilon D\tau_0^{11}(a\tau_0^{22}+\frac{\nu}{ReE})+ \nonumber\\
& a\varepsilon \tau_0^{11} D\tau_0^{22}+a D\tau_0^{11}(i\alpha(U-c)+\frac{\nu}{ReE}+\varepsilon(\tau_0^{11}+2\tau_0^{22})) + (a\tau_0^{11}+\frac{\nu}{ReE})(i\alpha(DU)+\varepsilon(D\tau_0^{11} \nonumber\\
& +2D\tau_0^{22}))+(\varepsilon \tau_0^{11}(1-a)+(i\alpha(U-c)+\frac{\nu}{ReE}+\varepsilon(\tau_0^{11}\!\!+\!\!2\tau_0^{22}))(1\!\!+\!\!a))(\varepsilon(D^2\tau_0^{11})(a\tau_0^{22}\!\!+\!\!\frac{\nu}{ReE}) \nonumber\\
& +2a\varepsilon(D\tau_0^{11})(D\tau_0^{22})+a\varepsilon\tau_0^{11}(D^2\tau_0^{22})+(aD^2\tau_0^{11})(i\alpha(U-c)+\frac{\nu}{ReE}+\varepsilon(\tau_0^{11}+2\tau_0^{22})) + 2aD\tau_0^{11} \nonumber\\
& (i\alpha(DU)+\varepsilon(D\tau_0^{11}+2D\tau_0^{22}))+(a\tau_0^{11}+\frac{\nu}{ReE})(i\alpha(D^2U)+\varepsilon(D^2\tau_0^{11}+2D^2\tau_0^{22})) - (2\varepsilon D^2\tau_0^{11}+ \nonumber\\ 
& (i\alpha(D^2U)+2\varepsilon D^2\tau_0^{22})(1+a))(\varepsilon \tau_0^{11})(a\tau_0^{22} + \frac{\nu}{ReE}) + (a\tau_0^{11} + \frac{\nu}{ReE})(i\alpha(U - c) + \frac{\nu}{ReE} + \varepsilon(\tau_0^{11} \nonumber\\
& + 2\tau_0^{22}))) - (2\varepsilon D\tau_0^{11}+(i\alpha(DU)+2\varepsilon D\tau_0^{22})(1+a))(\varepsilon D\tau_0^{11})(a\tau_0^{22} + \frac{\nu}{ReE}) +a\varepsilon\tau_0^{11} D\tau_0^{22}+ \nonumber\\
& a D\tau_0^{11}(i\alpha(U - c) + \frac{\nu}{ReE} + \varepsilon(\tau_0^{11} + 2\tau_0^{22}))+(a\tau_0^{11} + \frac{\nu}{ReE})(i\alpha(DU) + \varepsilon(D\tau_0^{11} + 2D\tau_0^{22})))) - \nonumber\\
& 2(2\varepsilon D\tau_0^{11}+(i\alpha(DU)+2\varepsilon D\tau_0^{22})(1+a))(\varepsilon D\tau_0^{11}(a\tau_0^{22}+\frac{\nu}{ReE})+a\varepsilon\tau_0^{11} D\tau_0^{22} + aD\tau_0^{11}(i\alpha(U-c) \nonumber\\
& + \frac{\nu}{ReE}+ \varepsilon(\tau_0^{11}+2\tau_0^{22}))+(a\tau_0^{11}+\frac{\nu}{ReE})(i\alpha(DU)+\varepsilon(D\tau_0^{11} + 2D\tau_0^{22}))-(2\varepsilon D\tau_0^{11} + ( i\alpha DU  \nonumber\\
& \left. + 2\varepsilon D\tau_0^{22})(1+a))(\varepsilon \tau_0^{11}(a\tau_0^{22}+\frac{\nu}{ReE}) + (a\tau_0^{11} + \frac{\nu}{ReE})(i\alpha(U-c) + \frac{\nu}{ReE}+ \varepsilon(\tau_0^{11}+2\tau_0^{22})))) \right]. 
\end{align}

\noindent The components of the base state extra elastic stress tensor, $\tau_0$, is given by
\begin{align}
\tau_0^{11} &= \frac{1}{24\varepsilon^2a^2(\frac{ReE}{\nu})^4M_1^\frac{1}{3}}\!\!\!\left[\!\!2(2)^\frac{1}{3}\!\!\varepsilon^2(\frac{ReE}{\nu})^6(1\!\!+\!\!a)^4\!\!\!-\!\!4(2)^\frac{1}{3}\varepsilon^2(\frac{ReE}{\nu})^6(1\!\!+\!\!a)^2(1\!\!-\!\!a^2)\!\!+\!\!2(2)^\frac{1}{3}\varepsilon^2(\frac{ReE}{\nu})^6(1\!\!-\!\!a^2)^2 \right. \nonumber\\
&-\!\!6(2)^\frac{1}{3}\varepsilon^2(\frac{ReE}{\nu})^8(1\!\!+\!\!a)^4(1\!\!-\!\!a^2)U^2\!\!+\!\!12(2)^\frac{1}{3}\varepsilon^2(\frac{ReE}{\nu})^8(1\!\!+\!\!a)^2(1\!\!-\!\!a^2)^2U^2 \!\!-\!\! 6(2)^\frac{1}{3}\varepsilon^2(\frac{ReE}{\nu})^8(1\!\!-\!\!a^2)^3U^2 \nonumber\\
& \left. -8a\varepsilon(1+a)(\frac{ReE}{\nu})^3M_1^{\frac{1}{3}} + (2M_1)^{\frac{2}{3}}\right], \nonumber \\
\tau_0^{22} &= - \frac{(1-a)}{(1+a)}\tau_0^{11}, \nonumber \\
\tau_0^{12} &= \frac{DU(1-(1-a)\tau_0^{11}(\frac{ReE}{\nu}))}{(1+\varepsilon    (\frac{ReE}{\nu})(\tau_0^{11}+\tau_0^{22}))},
\label{eqn:baseStress}
\end{align}
\noindent where the coefficient $M_1$ in the base stress, is given by
\begin{align}
M_1 &= 16\varepsilon^3(\frac{ReE}{\nu})^9a^3(1\!\!+\!\!a)^3\!\!\!\left[ \!\!(27a\varepsilon+9(1\!\!-\!\!a^2))(\frac{ReE}{\nu})^2U^2\!\!+\!\!1\!\!+\!\!3(3)^\frac{1}{2}(\frac{ReE}{\nu})(27\varepsilon^2(\frac{ReE}{\nu})^2a^2U^4\!\!+\!\!(1\!\!-\!\!a^2) \right. \nonumber\\
& \left. (U+(\frac{ReE}{\nu})^2(1-a^2)U^3)^2+2a\varepsilon(U^2+9(\frac{ReE}{\nu})^2(1-a^2)U^4))^\frac{1}{2}\right].
\end{align}

\bibliographystyle{unsrt}
\bibliography{PoF}

\end{document}